\documentclass[twocolumn,trackchanges]{aastex63}

\usepackage{enumitem}
\usepackage{CJKutf8}
\newcommand{\cntext}[1]{\begin{CJK}{UTF8}{gbsn}#1\end{CJK}\kern-1ex}

\submitjournal{ApJ}
\shorttitle{An \textit{in situ} Type III Radio Burst Event}
\shortauthors{Wang M. et al.}
\usepackage{amsmath}
\graphicspath{{./}{figures/}}

\begin{document}

\title{The Solar Origin of an \textit{In Situ} Type III Radio Burst Event}

\correspondingauthor{Bin Chen}
\email{bin.chen@njit.edu}

\author[0000-0002-2633-3562]{Meiqi Wang (\cntext{王美祺})}
\affiliation{Center for Solar-Terrestrial Research, New Jersey Institute of Technology, Newark, NJ 07102,  USA}

\author[0000-0002-0660-3350]{Bin Chen (\cntext{陈彬})}
\affiliation{Center for Solar-Terrestrial Research, New Jersey Institute of Technology, Newark, NJ 07102,  USA}

\author[0000-0003-2872-2614]{Sijie Yu (\cntext{余思捷})}
\affiliation{Center for Solar-Terrestrial Research, New Jersey Institute of Technology, Newark, NJ 07102,  USA}

\author[0000-0003-2520-8396]{Dale E. Gary}
\affiliation{Center for Solar-Terrestrial Research, New Jersey Institute of Technology, Newark, NJ 07102,  USA}

\author[0000-0002-5865-7924]{Jeongwoo Lee}
\affiliation{Center for Solar-Terrestrial Research, New Jersey Institute of Technology, Newark, NJ 07102,  USA}
\affiliation{Institute for Space Weather Sciences, New Jersey Institute of Technology, Newark, NJ 07102, USA}

\author[0000-0002-5233-565X]{Haimin Wang}
\affiliation{Center for Solar-Terrestrial Research, New Jersey Institute of Technology, Newark, NJ 07102,  USA}

\author[0000-0002-0978-8127]{Christina Cohen}
\affiliation{Space Radiation Laboratory, California Institute of Technology, Pasadena, CA 91125, USA}

\begin{abstract}

Solar type III radio bursts are generated by beams of energetic electrons that travel along open magnetic field lines through the corona and into interplanetary space. However, understanding the source of these electrons and how they escape into interplanetary space remains an outstanding topic. Here we report multi-instrument, multi-perspective observations of an interplanetary type III radio burst event shortly after the second perihelion of the Parker Solar Probe (PSP). This event was associated with a solar jet that produced an impulsive microwave burst event recorded by the Expanded Owens Valley Solar Array (EOVSA). 
The type III burst event also coincided with the detection of enhanced \textit{in situ} energetic electrons recorded by both PSP at 0.37 AU and WIND at 1 AU, which were located very closely on the Parker spiral longitudinally. 
The close timing association and magnetic connectivity suggest that the \textit{in situ} energetic electrons originated from the jet's magnetic reconnection region. 
Intriguingly, microwave imaging spectroscopy results suggest that the escaping energetic electrons were injected into a large opening angle of about 90$^{\circ}$, which is at least nine times broader than the apparent width of the jet spire.
Our findings provide an interpretation for the previously reported, longitudinally broad spatial distribution of flare locations associated with prompt energetic electron events and have important implications for understanding the origin and distribution of energetic electrons in the interplanetary space. 

\end{abstract}


\section{Introduction} \label{sec:intro}
 
Solar type III radio bursts are one of the most commonly observed types of radio bursts from the Sun. In the radio dynamic spectrum, they feature a rapid drift from high to low frequencies \citep{Wild1950}. Owing to their fast frequency drift rates, type III bursts are believed to be associated with fast electron beams propagating in the solar corona and interplanetary space at a fraction of the speed of light ($\sim$0.1--0.5 c). The widely adopted emission mechanism of type III bursts, first proposed by \citet{Ginzburg1958} and further developed by many researchers over decades, involves the generation of a bump-on-tail distribution by a propagating fast electron beam leading to nonlinear growth of Langmuir waves via the two-stream instability.  These Langmuir waves are subsequently converted to transverse electromagnetic waves near the local plasma frequency or its harmonic \citep[see, e.g.,][for a review]{Reid2014}. Theoretical and modeling studies have suggested that such a bump-on-tail instability can naturally develop when high-speed electron beams are traveling along magnetic field lines, leading to a positive gradient in the electron velocity distribution ($\partial f/\partial v > 0$) around the ``bump'' at the electron beam velocity $v_b$ \citep{Melrose1990,Reid2011}. The presence of the bump-on-tail feature in association with type III radio bursts has been supported by multiple \textit{in situ} observations \citep{Gurnett1975, Lin1985, Ergun1998}. 

Since \citet{vanAllen1965}, several analyses of solar energetic electron (SEE) events with \textit{in situ} observations have been reported \citep{Reames1990,Lin1996,Krucker1999, Krucker2007, Simnett2002}. \citet{wang2012} reported that almost all the observed \textit{in situ} SEE events are associated with type III radio bursts. For the energetic electron beams to enter interplanetary space to produce the low-frequency type III bursts $\sim$10 kHz to MHz and \textit{in situ} energetic electron events, they need to have access to open field lines. Solar jets, which manifest in extreme ultraviolet (EUV) and X-ray images as collimated material ejecting along apparently open field lines, provide evidence of the existence of such a magnetic topology. Indeed, it has been reported that some type-III-burst-producing \textit{in situ} SEE events are closely associated with jets (observed in EUV and/or X-ray) \citep{Aurass1994,Raulin1996,Klassen2011}. For example, by combining \textit{in situ} electron measurements made by the three-dimensional Plasma and Energetic Particles (3DP) experiment onboard the WIND spacecraft and remote-sensing hard X-ray (HXR) observations from the Reuven Ramaty High Energy Solar Spectroscopic Imager (RHESSI), \citet{Krucker2011} explored the source region of supra-thermal electron beams and suggested that they are associated with interchange reconnection occurring between open and closed magnetic field lines.

Although interplanetary type III radio bursts have been frequently reported to show a close temporal association with SEEs \citep{Lin1973, Cane2003a, wang2016, Gomez-Herrero2021}, identifying the source region of the type-III-burst-emitting electron beams remains challenging. Many studies utilize a magnetic field model to investigate the connection between the \textit{in situ} SEE measurements and features on the solar surface. 
Meanwhile, the source region of the SEEs has also been located by comparing the release time inferred from a velocity dispersion analysis or assuming the electrons’ scatter-free propagation along the Archimedean interplanetary magnetic field line to the timing of the radio, EUV, or X-ray emission signatures obtained via remote sensing observations \citep{Krucker1999, Haggerty2002}.

More recently, radio imaging spectroscopy has started to play a more and more important role in tracing the solar origin of energetic particles. A number of studies have analyzed the source of type III radio bursts associated with jets using radio imaging spectroscopy \citep{cairns2009,mccauley2017,mulay2019,chen2013,chen2018}. Particularly, radio imaging spectroscopy results in the hundreds of MHz to GHz range (or decimetric wavelengths) have indicated that the decimetric type III bursts are emitted by electron beams originating from the core region of the jets in the low corona, presumably close to the magnetic reconnection site. For example, \citet{Carley2016} have identified different acceleration sites of electrons associated with an erupting jet and flux rope by imaging radio sources at frequencies ranging from 150--445 MHz observed by Nan\c{c}ay Radioheliograph. In \citet{chen2018}, using the observations from the upgraded Karl G. Jansky Very Large Array (VLA) in the 1--2 GHz L band, they have derived detailed trajectories of type III-emitting electron beams and interpret each beam-diverging site as a reconnection null point.  In addition, for low-frequency interplanetary type III radio bursts which are not accessible by ground-based radio instruments (as they occur at frequencies below the ionospheric cutoff at $\sim$20 MHz), techniques such as radio triangulation \citep{Fainberg1972,Gurnett1978,reiner1988} and ray-tracing \citep{Thejappa2008,Thejappa2010,Musset2021} have been used to locate their source locations in interplanetary space. 

Thanks to the operation of the Expanded Owens Valley Solar Array (EOVSA; \citealt{Gary2018}), microwave imaging spectroscopy observations of the full solar disk has been available from 1 to 18 GHz in 451 spectral channels with 1 s time resolution. The operation of EOVSA provides means of combining imaging spectroscopy observations of microwave emission from nonthermal electrons in the low corona with \textit{in situ} observations from Parker Solar Probe (PSP), STEREO-A, and WIND, offering an opportunity to form a more comprehensive picture of where these energetic electrons are accelerated and how they are transported from the low corona into interplanetary space. 

Here we report an interplanetary type III radio burst event that was observed jointly by PSP, STEREO-A, and WIND, with counterparts in higher radio frequencies observed by e-Callisto and EOVSA. In Section \ref{sec:Multiwavelength observation}, we present the multi-instrument, wide-band dynamic spectroscopy observations of this interplanetary type III burst event. In Section \ref{sec:overview}, we give the overview of the jet and the active region. In Section \ref{sec:freq_drift} and \ref{sec:directivity}, we derive the propagation and directivity of the interplanetary type III burst event from near the solar surface to interplanetary space. In Section \ref{sec:in situ}, we present \textit{in situ} measurements made by the PSP and WIND, which suggest that the type-III-burst-emitting electron beams have reached the two spacecraft located at 0.37 AU and 1 AU, respectively. In Section \ref{sec:Magnetic connectivity}, we discuss the magnetic connectivity between the PSP and the jet event supported by a potential-field source-surface (PFSS) magnetic field model. Then, we perform imaging spectroscopy of the microwave counterpart of the interplanetary type III radio burst by using EOVSA data and identify the origin of the source electrons as those accelerated in the jet event. Finally, in Section~\ref{sec:Summary and Discussion}, we summarize our main results and discuss their implications.

\section{Observations} \label{sec:Multiwavelength observation}
\subsection{Event Overview} \label{sec:overview}

\begin{figure*}
    \center
    \includegraphics[width=0.7\textwidth]{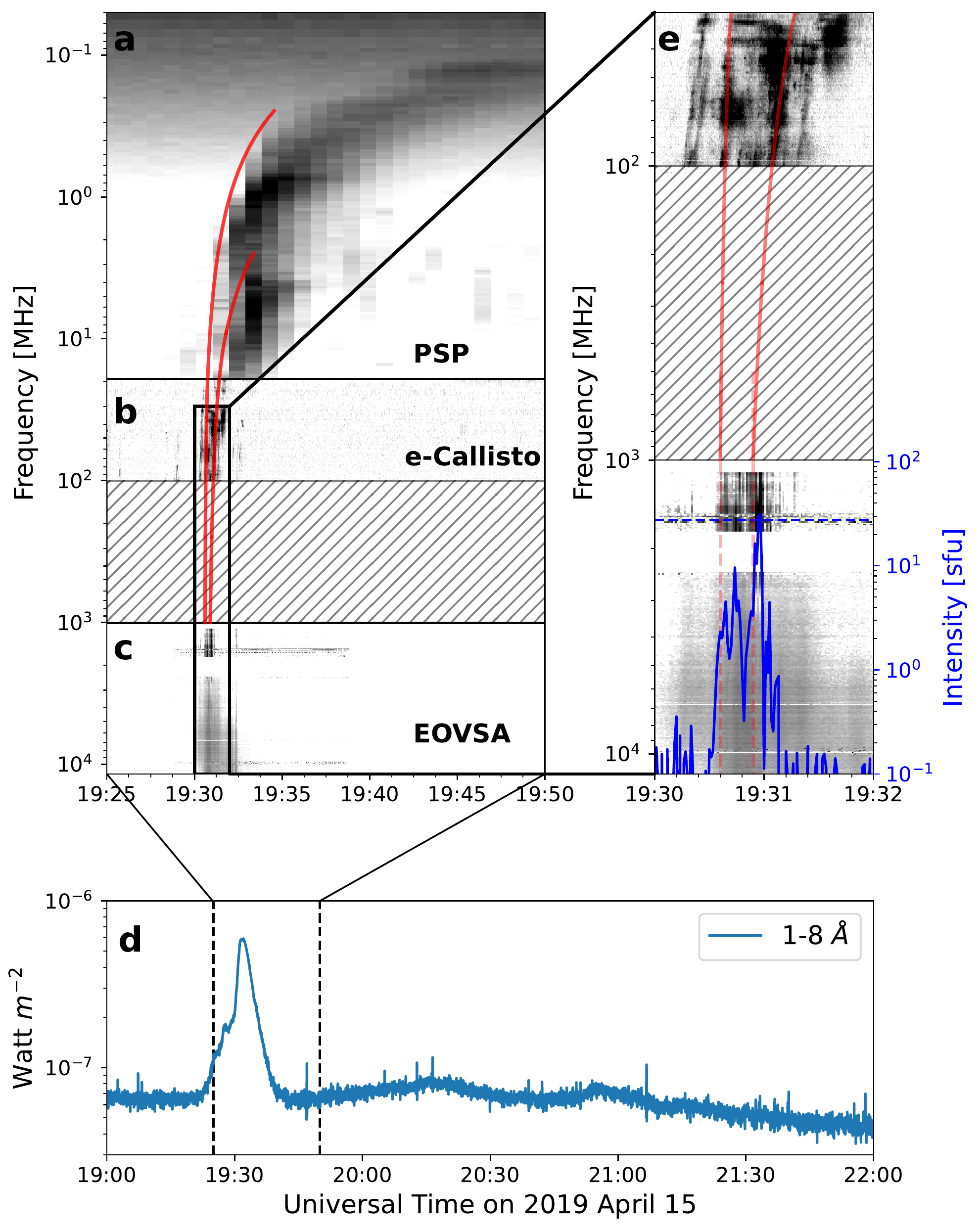}
    \caption{(a)--(c): Composite radio dynamic spectra from 50 kHz to 18 GHz featuring the interplanetary type III radio burst event observed by PSP/FIELDS (0.05--19 MHz), e-Callisto/Greenland (19--100 MHz), and EOVSA (1--18 GHz) during a GOES B3.4-class microflare on 2019 April 15 (d). An enlarged view of the high-frequency component of the burst event is shown in (e). Red curves show the frequency-drift fit for two individual type III bursts that can be traced to microwave frequencies. The blue curve represents the EOVSA 1.6 GHz light curve. The selected frequency is indicated by a blue horizontal dashed line.}
    \label{fig:overview}
\end{figure*}

The interplanetary type III radio burst event under study was recorded by three well-separated spacecraft as well as multiple ground-based radio facilities on 2019 April 15 around 19:30 UT, which was during the outbound phase of PSP's second perihelion. Figure \ref{fig:overview} shows the overall dynamic spectrum obtained by PSP/FIELDS/RFS \citep{field, Pulupa2017} at 50 kHz--20 MHz (panel (a)), e-Callisto/Greenland at 10 MHz--100 MHz (panel (b)), and EOVSA 1--18 GHz (panel (c)).

\begin{figure*}[!ht]
    \center
    \includegraphics[width=1.0\textwidth]{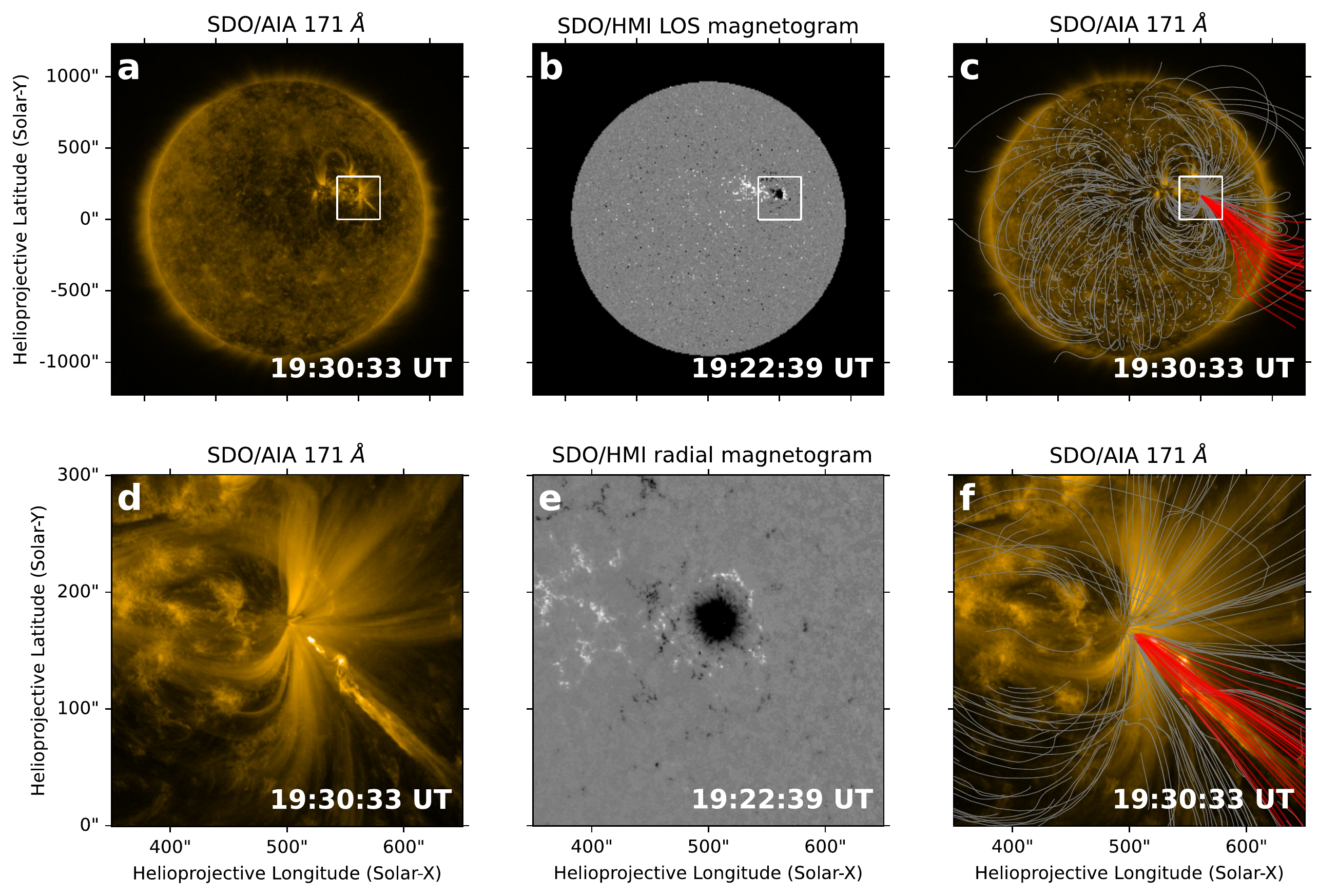}
    \caption{(a) SDO/AIA 171 \AA\ full disk image at 19:33 UT on 2019 April 15 around the peak of the B3.4 flare. (b) SDO/HMI full-disk line-of-sight magnetogram at 19:22 UT. 
    (c) Selected magnetic field lines from the PFSS extrapolation model. Red curves show the open field lines rooted at the leading sunspot with a negative magnetic polarity. (d)--(f) Detailed view of the region of interest in (a)--(c) (white box). Note that in panel (e), different from panel (b), the radial magnetogram is shown instead to mitigate projection effects. The transformation is based on the full vector magnetogram measurements of SDO/HMI.}
    \label{fig:jet_overview}
\end{figure*}

This interplanetary type III radio burst event coincided with a GOES soft X-ray (SXR) B3.4-class microflare, which started at $\sim$19:22 UT and peaked at 19:31 UT according to the 1--8 \AA\ SXR light curve shown in Figure \ref{fig:overview}(d). The microflare was the only flare event during the day. It was associated with a solar jet located in the periphery of AR 12738 featuring a long ($>$115 Mm) and collimated jet spire observed by multiple EUV channels of the Atmospheric Imaging Assembly \citep[AIA;][]{AIA} onboard the Solar Dynamics Observatory \citep[SDO;][]{SDO}. Figure \ref{fig:jet_overview}(a) and (d) show a full-disk view and a more detailed view of the jet event in AIA 171 \AA, respectively.

At the time of the interplanetary type III burst event, AR 12738 was located at a latitude of $6^{\circ}$ north and a longitude of $32^{\circ}$ west in heliographic coordinates. The active region consists of an $\alpha$-type sunspot (with a unipolar magnetic field configuration) surrounded by weak magnetic structures with mixed polarities. The full-disk line-of-sight (LOS) magnetogram is shown in Figure \ref{fig:jet_overview}(b), which is obtained by the Helioseismic and Magnetic Imager \citep[HMI;][]{HMI} onboard SDO. An enlarged view of the AR in EUV 171 \AA\ and the HMI radial magnetogram (calculated from components of the HMI vector magnetogram) is shown in Figures \ref{fig:jet_overview}(d) and (e). The jet is clearly visible at the west edge of the negative polarity sunspot and erupts in the southwest direction.

Selected field lines from the full-disk PFSS magnetic model, obtained by using the python package \texttt{pfsspy}; \citealt{Stansby2020}, are shown as gray lines in Figure \ref{fig:jet_overview}(c). The PFSS model is calculated using the HMI synoptic radial magnetogram for Carrington rotation 2216 (based on HMI LOS magnetograms collected over a 27-day solar rotation) as the input, which was then down-sampled to a size 360 pix $\times$ 180 pix. The source surface height is set at 2.5 $R_{\odot}$.
An enlarged view is shown in Figure~\ref{fig:jet_overview}(f), with PFSS extrapolated field lines overlaid on the SDO/AIA 171 \AA\ image. A few selected open field lines originating from the negative spot are shown as red curves which, as expected, coincide well with the EUV jet spire. We note, however, the highly-collimated EUV jet spire is much narrower than the full extent of the open field lines seen in the PFSS results. 

The event occurred during the period of a deep solar minimum. The background GOES 1--8 \AA\ SXR flux level remained very low and below the B level throughout the day. In addition, AR 12738 was the sole active region visible on disk over the period of 2019 April 10--17. 
The relatively quiet period, low background, and the presence of a single, dominant active region make it convenient to study the connections between the interplanetary type III radio bursts and \textit{in situ} particle events, as well as identify their source region. AR~12738 was the source of other published events.
For example, \citet{Badman2022} reported an interplanetary type III radio burst event on 2019 April 9 jointly observed by PSP/FIELDS, STEREO/WAVES, as well as the ground-based Low Frequency Array (LoFAR), and traced the source electrons to AR 12738. \citet{Wiedenbeck2020} reported $^3$He-rich SEPs events recorded by PSP/IS$\odot$IS and ACE from 2019 April 20--21 and suggested that they all originated from AR 12738. Interestingly, on 2019 April 4, one day prior to its perihelion, PSP observed a small SEP event. The source of this event was identified, by \citet{Leske2020}, as the same active region but was located nearly 80$^{\circ}$ east of the nominal PSP magnetic footpoint, implying that the open magnetic field lines along which the particles escape may extend over a wide longitudinal range.

\subsection{Interplanetary type III Radio Burst Event: Spectro-Temporal Properties} \label{sec:freq_drift}

\begin{figure*}
    \center
    \includegraphics[width=1\textwidth]{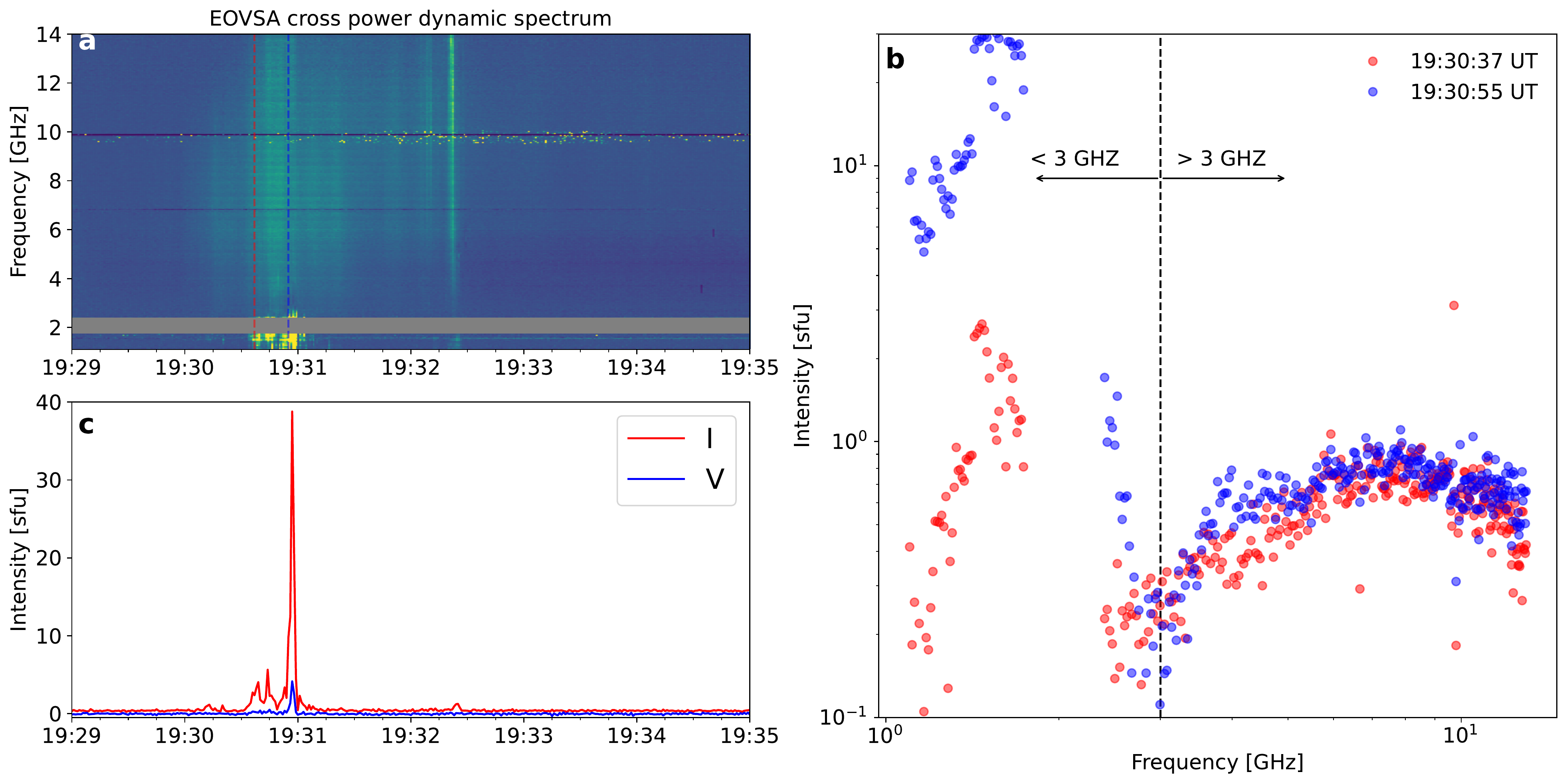}
    \caption{(a) EOVSA cross-power dynamic spectrum from 1--14 GHz. The red and blue dashed lines correspond to the start times of the two interplanetary type III burst branches in Figure~\ref{fig:overview}. (b) The spectra extracted from the two selected times in (a). (c): Stokes I (red) and V (blue) light curves at 1.5 GHz. At the peak time of 19:30:55 UT, the maximum degree of circular polarization is around 9$\%$.}
    \label{fig:crosspower}
\end{figure*}

In the composite dynamic spectrum from 50 kHz to 18 GHz shown in Figure~\ref{fig:overview}, the PSP interplanetary type III burst event consists of at least two individual branches that continue to the metric-microwave range measured by e-Callisto and EOVSA. Their characteristic rapid frequency drift from high to low frequencies can be understood by the propagation of electron beams from the low corona to interplanetary space during which they encounter background plasma with a decreasing density \citep{Wild1950}. The relationship between the emission frequency $\nu$ (in Hz) and electron density $n_e$ (in cm$^{-3}$) is given by $\nu \approx s\nu_{pe} = 8980s\sqrt{n_e}$ Hz, where $\nu_{pe}$ is the plasma frequency and $s\approx1$ or 2 is the harmonic number for the fundamental or harmonic plasma radiation, respectively.

To model the frequency drift of the interplanetary type III radio bursts, we adopt an electron density model $n_e(r)$ by combining the Newkirk density model in the low corona ($r=1.02$--1.5 $R_\odot$; \citealt{Newkirk1967}) with a solar wind density model in the upper corona and interplanetary space ($r>1.5\ R_\odot$; \citealt{Saito1977}). The low corona and solar wind density models are scaled by a factor of 0.5 and 9, respectively, in order to ensure a smooth transition at $1.5\ R_\odot$ and match the \textit{in situ} measurement of the plasma density made by PSP/SWEAP at 79 $R_\odot$. The exact form of our adopted density model is

\begin{equation}
    n_e(r)=\begin{cases}
    2.1 \times 10 ^ 4 \times 10 ^ {4.32/r}&\text{(1.02 $<$ r $<$ 1.5)}\\
     & \\
    4.62 \times 10^5r^{-2} + 4.74\times 10^7r^{-3.3} \\+ 3.19\times10^7 r^{-5.8}&\text{(1.5 $\leq$ r $<$ 215)}
    \end{cases}
    \label{eq:density}
\end{equation}
where $r$ is the heliocentric distance in solar radii.

Assuming a constant electron beam speed $v_b$, the distance of the interplanetary type III source from the Sun is
\begin{equation}
r(t) = v_b (t-t_0) \label{distance},
\end{equation}
where $t_0$ is the release time of the electron beam. 
Therefore, by assuming fundamental plasma radiation at $\nu\approx 1.2\nu_{pe}$ (e.g., \citealt{Musset2021}), the frequency variation in heliocentric distance $\nu(r)$ can be, in turn, expressed as a function of time $t$. 

By varying the electron beam speed $v_b$ and the release time $t_0$, we fit the leading edge of the two interplanetary type III burst branches in the dynamic spectrum from 100 kHz to $\sim$100 MHz. For the high-frequency range ($\sim$100 MHz--1 GHz) where the density model is out of range,  we extrapolate the same frequency drift curves beyond $\sim$100 MHz. The best-fit results are shown as red curves in Figures~\ref{fig:overview}(a) and (b). 
The first frequency drift curve extends all the way from $<$2 GHz to 200 kHz. The second curve appears to stop at a relatively higher frequency ($\sim$3 MHz). The fitted electron beam speed $v_b$ values are 0.35c and 0.1c, respectively. 
The extrapolated start time of the second frequency-drift curve coincides very well with the onset of the strongest EOVSA 1--3 GHz microwave burst at 19:30:55 UT (blue curve in Figure~\ref{fig:overview}(e)). The start time of the first curve also seems to coincide with a small microwave peak at 19:30:37 UT, but the correspondence is relatively less clear.

The well-correlated timing of the interplanetary type III radio bursts and the EOVSA bursts in $\sim$1--3 GHz suggests that the latter may be the decimetric counterpart of the interplanetary type III bursts due to coherent plasma radiation. To further elucidate the coherent nature of the decimetric bursts, we show a detailed EOVSA Stokes I dynamic spectrum with full frequency resolution (451 frequency channels spread in 1--18 GHz) and time resolution (1 s) in Figure~\ref{fig:crosspower}(a). The dynamic spectrum is obtained by taking a median over the cross-power dynamic spectra of all baselines shorter than 1.55 km, which reveals the relatively weak microwave bursts. In Figure~\ref{fig:crosspower}(b), we show the cross-power spectra for two selected times that correspond to the starting time of the two interplanetary type III burst branches (indicated by the red and blue dashed lines in Figure~\ref{fig:crosspower}(a)). The spectra at high frequencies ($>$3 GHz) have a positive slope followed by a negative slope at above $\sim$8 GHz,  which is consistent with incoherent nonthermal gyrosynchrotron emission \citep[e.g.,][]{Dulk1985}. At the same time, the spectra at $<$3 GHz are notably different, with much greater intensity and many fine structures. The characteristics of the EOVSA 1--3 GHz bursts are highly consistent with decimetric type III bursts reported in the literature \citep[e.g.,][]{Young1961,Paesold2001}, although EOVSA's temporal resolution (1 s) does not allow us to further resolve their presumably sub-second fine structures as presented in the observations by \citet{chen2013,chen2018} using the Jansky VLA. 

To further elucidate the nature of the 1--3 GHz microwave bursts, we investigate the circular polarization of the bursts. Although EOVSA is equipped with broadband cross-dipole feeds to offer imaging spectropolarimetry capabilities, it has antennas with dissimilar mount types: four of them are equatorially mounted, while the others are azimuth-elevation mounted. The different antenna mount types have resulted in polarization mixing caused by the differential rotation of the feeds.  
To avoid the complexity, we derive the Stokes V dynamic spectra by only using the six baselines of the four equatorially mounted antennas. The Stokes V light curve at 1.5 GHz, representing the median value of the six baselines, is shown as the blue curve in Figure~\ref{fig:crosspower}(c). The polarization degree remains consistently low throughout the entire event. At 19:30:55 UT, the polarization degree reaches a maximum of 9\%. We conclude that the weak polarization degree of the 1--3 GHz microwave bursts is consistent with harmonic plasma radiation \citep[e.g.,][]{Dulk1980}. The positive Stokes V value indicates that the bursts are right-hand circularly polarized (RCP). Considering the dominant negative sunspot, the emission is likely polarized in the sense of ordinary mode, or $o$ mode.

\subsection{Interplanetary type III Radio Burst Event: Location and Directivity} \label{sec:directivity}

\begin{figure*}
    \center
    \includegraphics[width=0.8\textwidth]{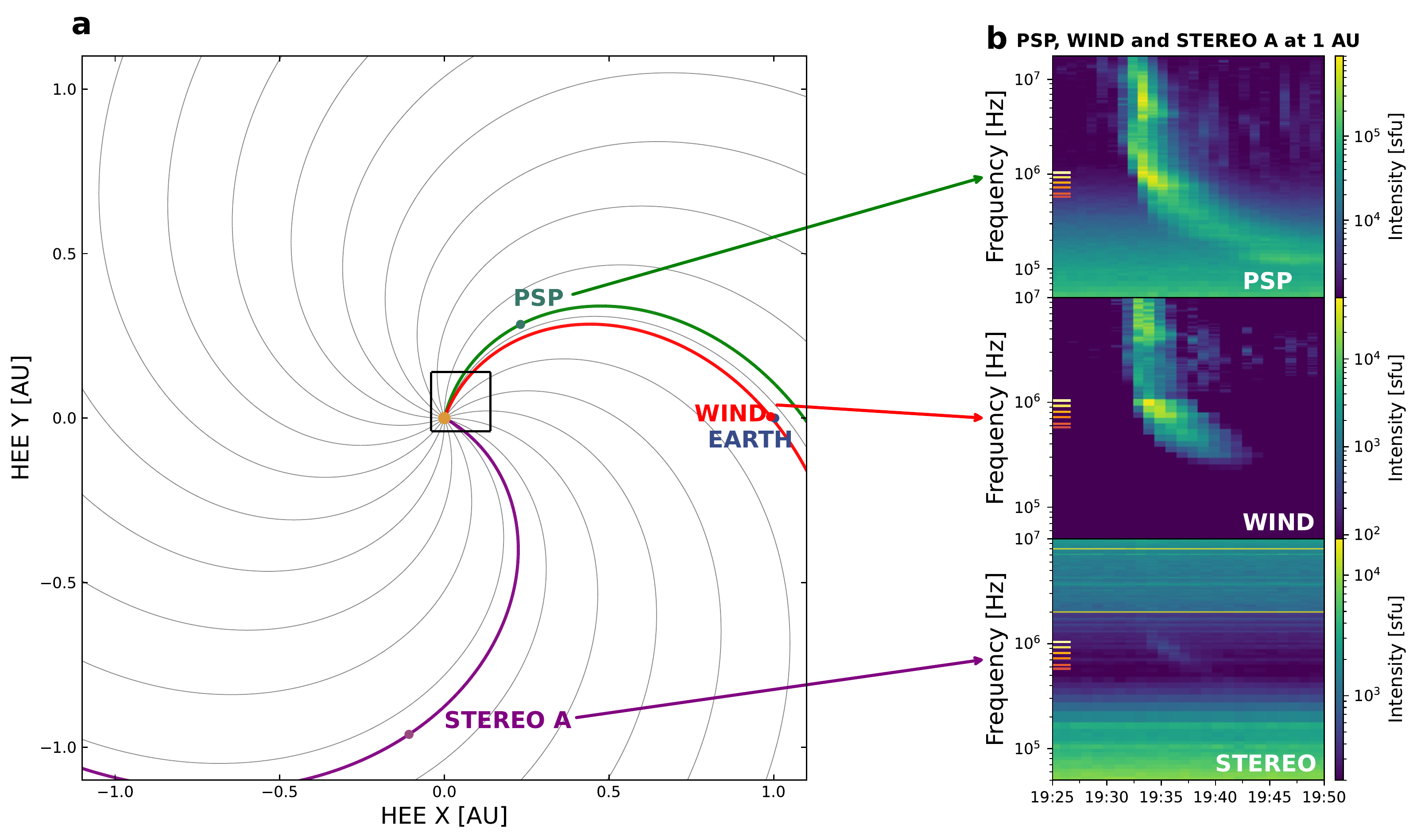}
    \caption{Multi-spacecraft observations of the interplanetary type III radio burst event. (a) Locations of PSP, WIND, and STEREO-A in the Heliocentric Earth Ecliptic (HEE) coordinate system. The gray curves in the background show the Parker spiral magnetic field lines in the ecliptic plane, equally spaced in longitude by 18 degrees. A solar wind speed of 350 km~s$^{-1}$ is assumed. Solid green, red, and purple curves are the magnetic field lines connected to PSP, Wind, and STEREO A. (b) Dynamic spectrum recorded by PSP/FIELDS, WIND/WAVES, and STEREO-A/WAVES. The light-travel delays and the intensity due to the different spacecraft positions have been corrected as if they are observed from 1 AU. The horizontal bars show the six selected frequencies (576, 625, 723, 815, 925, and 1040 kHz) used for the calculation of the interplanetary type III burst source position. }
    \label{fig: spacecraft location}
\end{figure*}

\begin{figure*}
    \center
    \includegraphics[width=0.8\textwidth]{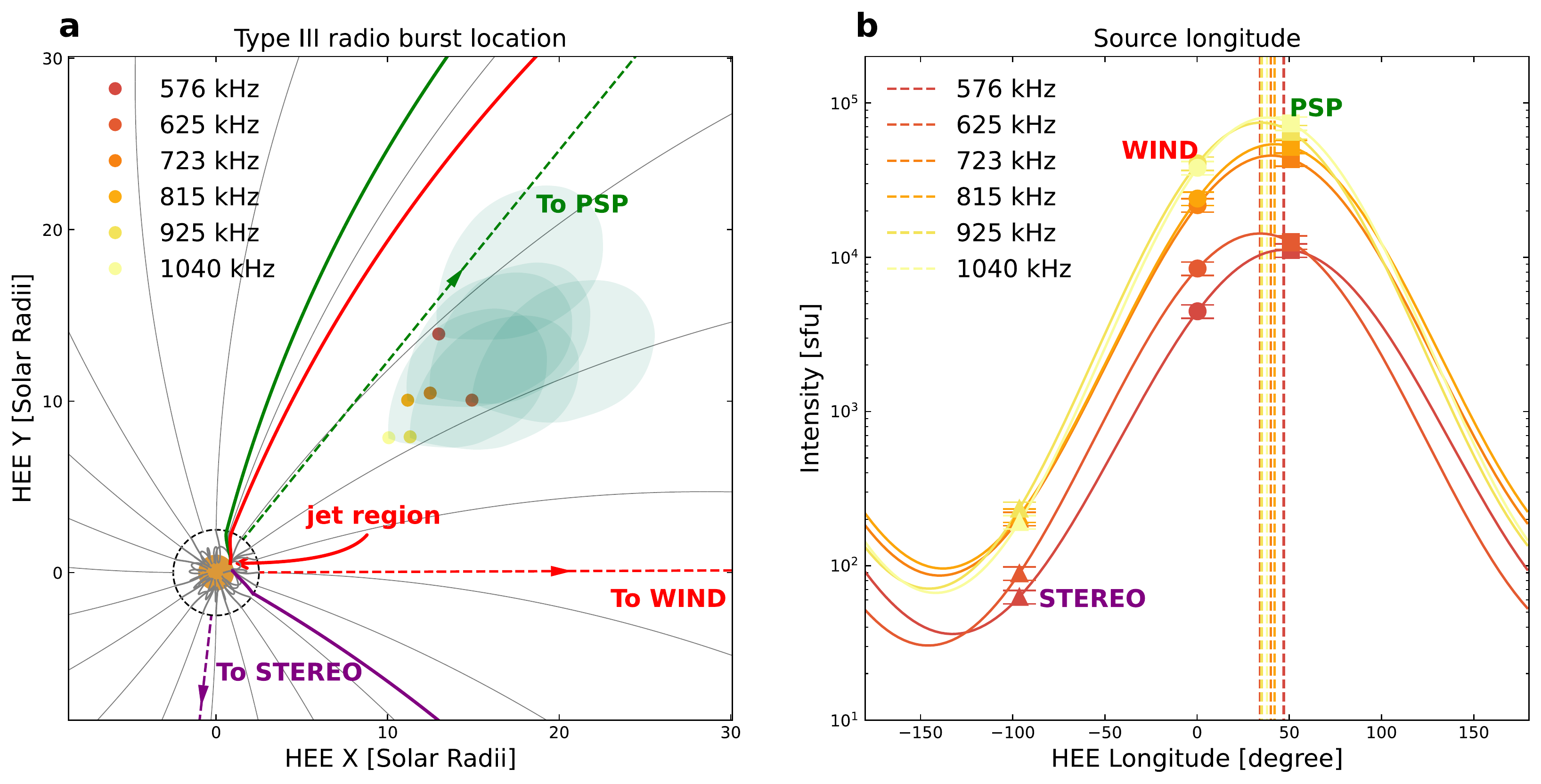}
    \caption{Locations of the interplanetary type III radio bursts. (a) Enlarged view of the inner heliosphere (corresponds to the black box in Figure~\ref{fig: spacecraft location}). Color circles show the derived source locations of the interplanetary type III bursts at six selected frequencies based on a joint analysis of observations made by all three spacecraft. The light green lobe at each burst location illustrates the derived emission pattern. The half-power beam width of each lobe represents the FWHM width of the estimated type III radio emission pattern. The solid red, green, and purple curves are the magnetic field lines connected to WIND, PSP, and STEREO-A, constructed by combining PFSS extrapolation and Parker spiral ballistic mapping results joining at a source surface of 2.5$R_{\odot}$ (dashed circle). WIND and PSP are magnetically connected to the jet region, while STEREO is not (see Section~\ref{sec:Magnetic connectivity}). Straight dashed lines are the line of sight directions of the three spacecraft. (b) Measured interplanetary type III burst intensities by the three spacecraft, scaled to 1 AU, at different HEE longitudes and the corresponding fit results of their angular emission patterns at the six selected frequencies. The vertical dashed lines mark the derived source longitudes at their respective peaks. A relative uncertainty of 10\% for the observed intensity is assumed, shown as the error bars.}
    \label{fig:type3_directivity}
\end{figure*}

This interplanetary type III radio burst event was recorded by three spacecraft---PSP/FIELDS, WIND/WAVES, and STEREO-A/WAVES observing the Sun from different viewing perspectives and heliocentric distances. The locations of the three spacecraft in the Heliocentric Earth Ecliptic (HEE) coordinate system are shown in Figure~\ref{fig: spacecraft location}(a). At the time of the event, all three spacecraft were located within (WIND and STEREO-A) or very close to (PSP) the ecliptic plane. Both WIND/WAVES and STEREO-A were located at a radial distance of $\sim$1 AU, while PSP was much closer at 0.37 AU away from the Sun. 
Figure~\ref{fig: spacecraft location}(b) shows the radio dynamic spectra of the interplanetary type III burst event observed by PSP/FIELDS, WIND/WAVES, and STEREO-A/WAVES, respectively, in 10 kHz--20 MHz. The data in the units of measured voltage power ($\mathrm{V}^2\ \mathrm{Hz}^{-1}$) have been calibrated to physical values in solar flux units (sfu) \citep{Dulk2001, Hillan2010,Eastwood2009,Pulupa2017}. The widely-separated, multi-vantage-point observations from the three spacecraft makes it possible to constrain the location and directivity of the interplanetary type III radio burst as it propagates in interplanetary space by using their relative intensity and timing (see, e.g., recent studies by \citealt{Musset2021,Badman2022}). 
 
Here we follow the method used in \citet{Musset2021} to perform the position and directivity analysis. First, we assume that the angular emission pattern of the interplanetary type III bursts at a source location $\vec{r}$ follows the form:
\begin{equation}
    I(\vec{r},\mu)=C_0 \exp\left(-\frac{1-\mu}{\Delta\mu}\right), \label{eq:tp3_directivity}
\end{equation}
where $\mu=\cos(\varphi - {\varphi}_0)$ describes the longitudinal angle of the emission pattern $\varphi$ with regard to the peak direction ${\varphi}_0$. $\Delta\mu$ is a parameter for the width of the emission pattern. Because the active region is located at a low solar latitude and all probes are located within or very close to the ecliptic plane, for simplicity, we only consider the longitudinal dependence of the radio emission pattern. 

The radio waves generated at the source location $\vec{r}$ then propagate through interplanetary space and reach the radio instruments onboard the three spacecraft.  Assuming free-space propagation of the radio waves from the source, the intensity of the observed radio intensity at the location of a spacecraft $i$ $\vec{r}_i^{obs}$ follows an inverse square law $I_i^{obs}\propto 1/(\vec{r}_i^{obs}-\vec{r})^{2}$. In addition, a spacecraft located at a larger distance from the source would have observed the burst with a larger time delay due to the longer light travel time. Such an effect is clearly shown when comparing the dynamic spectra obtained by, e.g., PSP and WIND. 

To correct for these effects associated with the different source--spacecraft distances, before we carry out the directivity analysis, we scale the radio intensities of all three spacecraft to what would have been measured if the spacecraft was positioned at 1 AU. We also shift all the spectra in time accordingly to compensate for the light travel time. We select six frequencies (576, 625, 723, 815, 925, and 1040 kHz) that are clearly observed by all three spacecraft. The scaled intensities at the six frequencies are shown in Figure~\ref{fig:type3_directivity}(b) as square, circle, and triangle symbols. The measurements made by the three spacecraft at each frequency are fitted using a least-square-based minimization technique, shown as color curves in Figure~\ref{fig:type3_directivity}(b). The dashed vertical lines correspond to the longitudes of the peak intensity $\varphi_0$ at the six selected frequencies, which range from 38 to 47 degrees. The fitted values of $\Delta\mu$ in Equation~\ref{eq:tp3_directivity} at different frequencies range from 0.28--0.35, corresponding to an average full width at half maximum (FWHM) angle of 73--81$^{\circ}$. Such an angular extent is consistent with those shown in \citet{Musset2021} for another interplanetary type III burst event. The radial distances of the type III burst source are calculated using the density model of Equation~\ref{eq:density}, again assuming fundamental plasma radiation emitted at $1.2\nu_{pe}$.

Figure~\ref{fig:type3_directivity}(a) shows the derived interplanetary type III burst source locations and the directions of their emission pattern at the six selected frequencies as color circles and light green lobes, respectively. 
The half-power beam widths of the lobes represent the FWHM width of each fitted angular emission pattern shown in Figure~\ref{fig:type3_directivity}(b). Type III emission theories suggest that the emission pattern of type III bursts resembles approximately a dipole or quadrupole shape and is generally aligned with the tangent of the magnetic field line along which the type-III-emitting electron beams are propagating \citep{Zheleznyakov1970,Thejappa2012,Thejappa2015}, although the exact source location and directivity pattern of the radio emission is strongly modulated by propagation effects in the inhomogeneous turbulent corona \citep[e.g.,][]{Kontar2019}. 
We conclude that the derived directivity of the interplanetary type III bursts is consistent with the scenario in which the type-III-emitting electron beams are directed toward the general direction of PSP and WIND. In our analysis, we simplify our approach by assuming that the interplanetary type III bursts are due to fundamental plasma radiation, which has a dipole-shaped emission pattern \citep{Zheleznyakov1970}. The situation would be different for harmonic radiation as the emission pattern would have a more complex quadrupole-type shape. We note that the assumption taken here is different from the decimetric bursts discussed in section \ref{sec:freq_drift}, which has been characterized as emission due to harmonic plasma radiation.

\subsection{Interplanetary type III Radio Burst Event: \textit{In Situ} Signatures at 0.37 AU and 1 AU} \label{sec:in situ}

\begin{figure*}
    \center
    \includegraphics[width=0.7\textwidth]{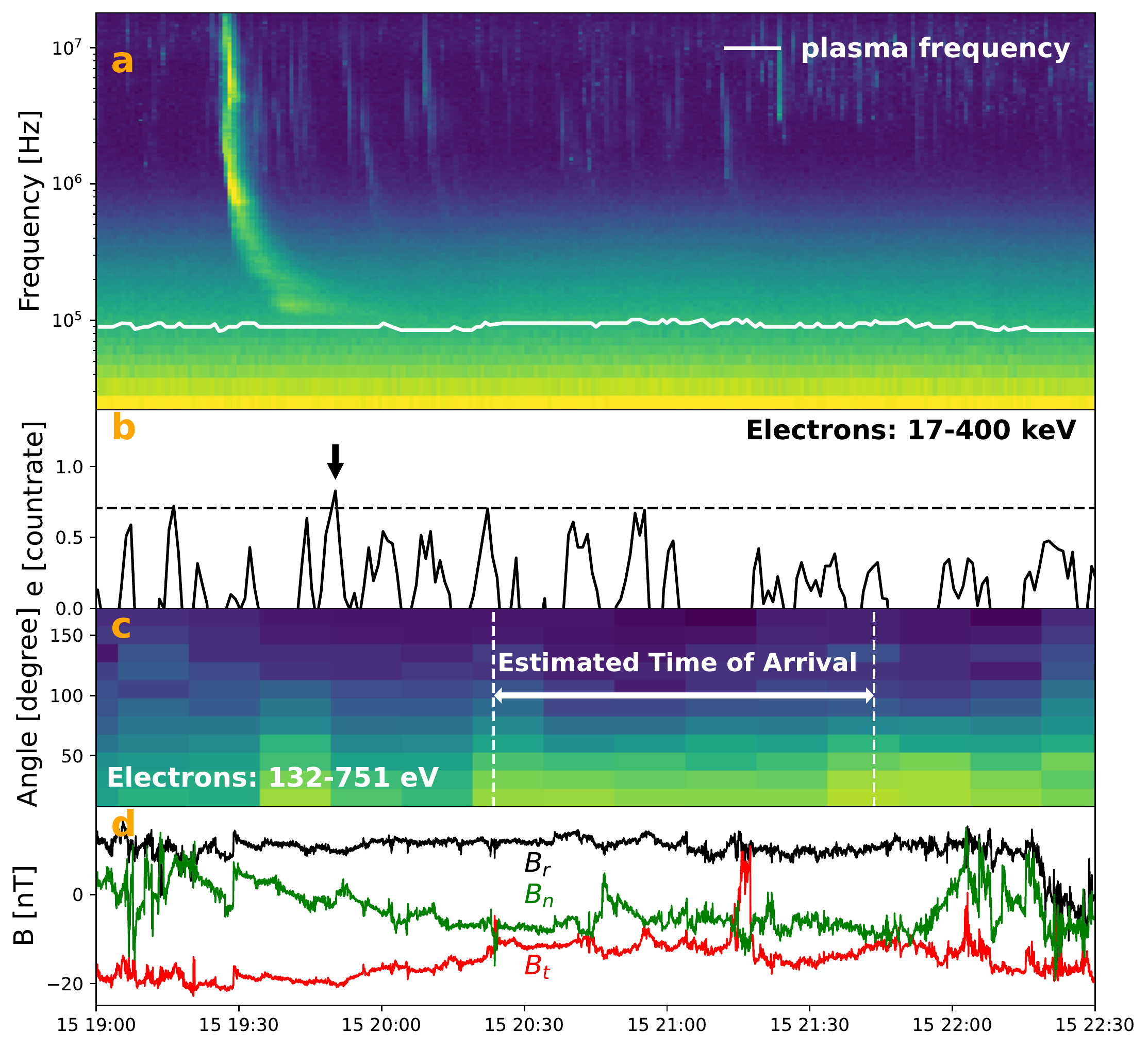}
    \caption{(a): Composite radio dynamic spectrum using data from PSP/FIELDS. The white curve shows where the local plasma frequency lies at the vicinity of the spacecraft. (b) Electron count rate from the EPI-Lo instrument onboard PSP/IS$\odot$IS. The data shown is background subtracted and smoothed using the Savitzky--Golay method. (c) Electron pitch angle distribution integrated over the energy range of 132--751 eV, obtained by the SPAN-E instrument onboard PSP/SWEAP. The two vertical dashed lines demarcate the estimated arrival times of the 132--751 eV electrons at the PSP spacecraft if we assume these electrons propagate freely along the Parker spiral. (d) PSP/FIELDS measurements of the magnetic field at the spacecraft. Three magnetic components in the radial--tangential--normal (RTN) coordinates are shown.}
    \label{fig: PSP observation}
\end{figure*}

In the previous subsection, the derived locations of the interplanetary type III sources and their emission patterns have implied that the type-III-burst-emitting electron beams may be propagating along magnetic field lines that are connected to PSP at 0.37 AU and WIND/Earth at 1 AU, which were located very closely on the Parker spiral longitudinally. In this subsection, we will discuss \textit{in situ} measurements made by PSP and WIND with a focus on signatures of the energetic electrons. 

First, we examine whether the interplanetary type III burst event indeed constitutes an \textit{in situ} event as seen by the PSP spacecraft. One of the main signatures of an \textit{in situ} type III burst event is the presence of a low-frequency cutoff near the local plasma frequency \citep[e.g.,][]{Dulk1998,Cane2003a}. As the type-III-burst-emitting electron beam reaches the vicinity of the spacecraft, the plasma emission is produced near the local plasma frequency. No emission at lower frequencies is allowed to propagate, forming a low-frequency cutoff.  Figure~\ref{fig: PSP observation}(a) shows the composite dynamic spectrum in 10 kHz--20 MHz obtained by PSP/FIELDS. It is evident that the interplanetary type III burst does show such a low-frequency cutoff near the local plasma frequency line (white curve) derived using the quasi-thermal noise data from PSP/FIELDS \citep{Moncuquet2020}. 
This observation implicates that the escaping electron beams have reached at least the same radial distance from the Sun as that of PSP.


In Figure \ref{fig: PSP observation}(b), we show the $>$17 keV energetic electron count rate recorded by PSP's Integrated Science Investigation of the Sun (IS$\odot$IS; \citealt{McComas2016}). The curve shown is obtained by the IS$\odot$IS Energetic Particle Instrument that measures the energetic particles with relatively lower energy (EPI-Lo). The data are first integrated within the energy range of 17--400 keV after performing background subtraction. Then, following the method in \citet{Mitchell2020}, a Savitzky--Golay smoothing technique with a 7-minute smoothing window is used to reduce the noise. The horizontal dashed line marks the 1$\sigma$ deviation above the mean. A small enhancement above the 1$\sigma$ level indicated by the black arrow in Figure~\ref{fig: PSP observation}(b) appears $\sim$20 minutes after the onset of interplanetary type III radio burst. However, owing to its low signal-to-noise ratio, we consider this enhancement at best a marginal detection. Nevertheless, we note that this enhancement occurs very close to the time when the low-frequency cutoff of the interplanetary type III radio burst is observed. Such a temporal coincidence is another possible indication of the arrival of the energetic electrons at the PSP spacecraft. 

\begin{figure*}
    \center
    \includegraphics[width=0.8\textwidth]{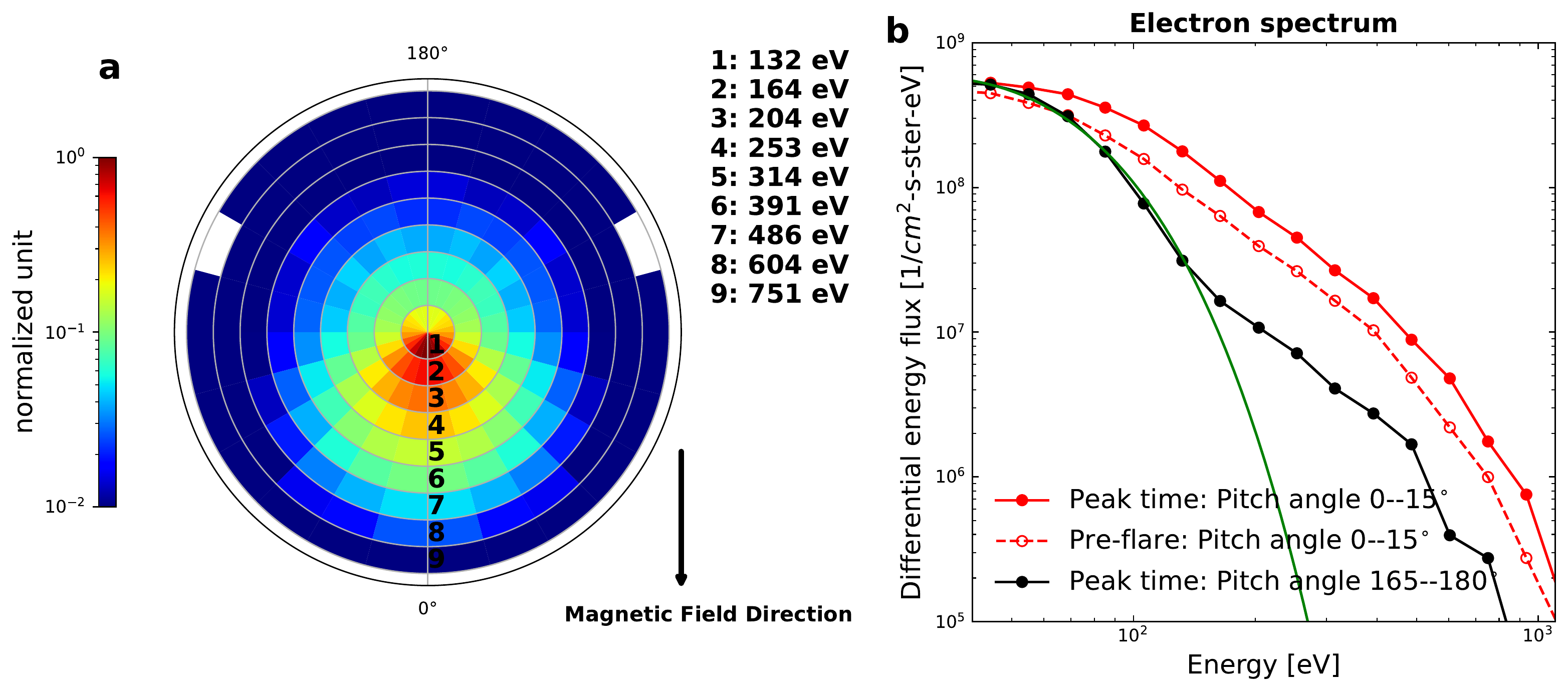}
    \caption{(a) Electron pitch-angle distribution in different energy bins from 132 eV to 751 eV as measured by PSP/SWEAP. The differential energy flux has been normalized to the maximum at each energy bin. Note that the distribution in 180$^{\circ}$--360$^{\circ}$ is mirrored from the measurements for 0$^{\circ}$--180$^{\circ}$. (b) Differential electron flux spectrum made at times during the peak (black and red solid curves) and just prior to the peak (dashed curve). An angle of zero degree is aligned with the direction of the magnetic vector. The green curve is the thermal distribution calculated with PSP/FIELD plasma measurements (with a temperature of $k_BT\approx 20\ \mathrm{eV}$).}
    \label{fig: electron pitch angle distribution}
\end{figure*}

\begin{figure}
    \center
    \includegraphics[width=0.5\textwidth]{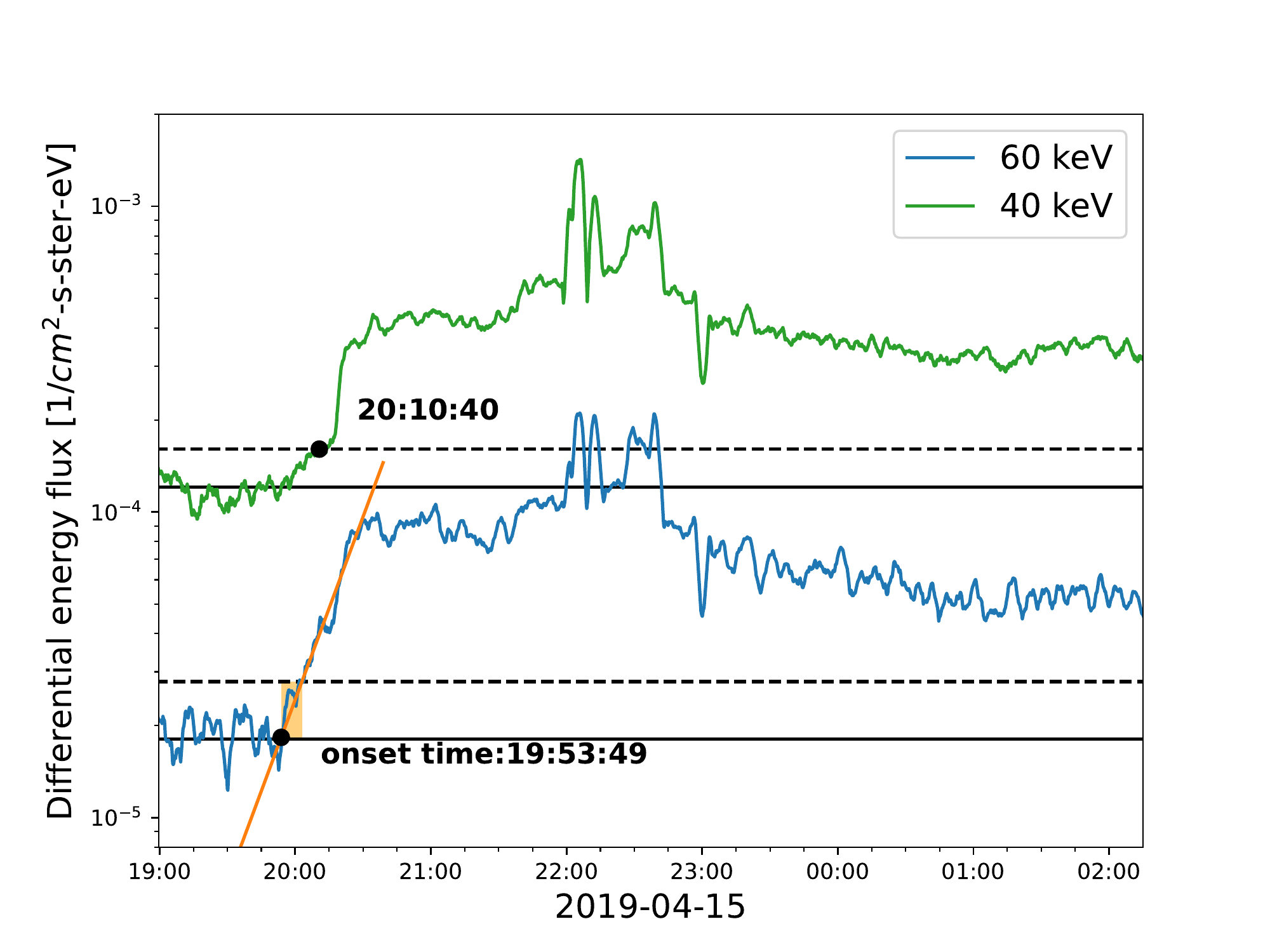}
    \caption{Time profiles of the \textit{in situ} energetic electron flux at 40 and 60 keV observed by Wind/3DP. The horizontal solid and dashed lines mark, respectively, the average pre-event background level and the $3\sigma$ level above the background. The orange rectangle indicates the range of the estimated onset time of the energetic electron event at 60 keV. The black circle indicates the estimated onset time at 40 keV.}
    \label{fig: wind observation}
\end{figure}

PSP is also equipped with electrostatic analyzers (referred to as Solar Probe ANalyzers, or SPANs) onboard the Solar Wind Electrons Alphas and Protons (SWEAP; \citealt{Kasper2016}) instrument suite that can measure the energy spectra and directivity of lower energy electrons ($\lesssim$1 keV). Figure~\ref{fig: PSP observation}(c) shows the evolution of the pitch angle distribution of the 132--751 eV electrons using SWEAP/SPAN-E data. The direction of the magnetic field is at the 0-degree pitch angle. Although the electrons are more concentrated along the direction of the magnetic field, representing the ubiquitous electron ``strahl'' population in the solar wind \citep{{Feldman1975, Rosenbauer1977, Pilipp1987}}, such a field-aligned concentration shows an evident enhancement after 20:20 UT. To further elucidate the pitch angle distribution of the electrons as a function of energy, we make a polar plot by averaging the SWEAP/SPAN-E data during the peak time between 20:19 UT and 21:18 UT, shown in Figure~\ref{fig: electron pitch angle distribution}(a). The different concentric rings in the polar plot correspond to nine energy channels ranging from 132 eV to 751 eV. The differential energy flux of each channel has been normalized to their respective maximum. The anti-sunward beam-like electron distribution can be seen clearly in all energy channels except for the highest energy channel (751 eV) where the signal is very weak. 
The pitch angle width at half maximum of the beam-like distribution decreases slightly from $65^{\circ}$ to $57^{\circ}$ in increasing energy. 

In Figure~\ref{fig: electron pitch angle distribution}(b), we show 1-D slices of the electron distribution made during the peak time for directions parallel (red solid curve) and anti-parallel (black solid) to the magnetic field. The thermal electron distribution at the respective time period is shown as the solid green curve, calculated using the thermal electron temperature derived from the plasma measurements made by PSP/FIELDS. In comparison, the red dashed curve in Figure~\ref{fig: electron pitch angle distribution}(b) shows the time just prior to the enhancement of electron flux (averaged from 18:05:00--19:15:00 UT). While a suprathermal electron population is present at all times (both prior to and during the peak), the electron flux is greater by a factor of 2 at all energies during the enhancement and is predominant along the magnetic field direction. 
The enhancement of the low-energy, $\sim$0.1--1 keV electron flux with a beam-like distribution closely resembles properties of the low-energy ``solar electron burst'' events reported in the literature \citep{Gosling2003, Gosling2004,deKoning2006}, which have been found to be strongly associated with the occurrence of interplanetary type III radio bursts that extended to the local plasma frequency. The low temporal resolution of our data (recorded 11 days after the perihelion) does not allow us to distinguish signatures of systematic time-of-the-flight delays in decreasing electron energy of the \textit{in situ} electrons as reported in, e.g., \citet{Gosling2003}. However, the timing of the enhancement is consistent with the period expected for the arrival of the $\sim$0.1--1 keV electrons (20:23--21:43 UT, indicated by the interval demarcated by the dashed vertical lines in Figure~\ref{fig: PSP observation}(c)), if we assume that they originate close to the solar surface at the start time of the interplanetary type III radio bursts and propagate along the Parker spiral freely (shown as the double-sided arrow in Figure~\ref{fig: PSP observation}(d)).

Now we turn to \textit{in situ} signatures of energetic electrons obtained at 1 AU using measurements made by the 3D Plasma and Energetic Particle Instrument onboard the WIND spacecraft (WIND/3DP; \citealt{Lin1995}). In Figure~\ref{fig: wind observation}, we show the time profile of the energetic electron flux from 19 UT to 24 UT in two energy bands at 40 keV and 60 keV. An enhancement is observed around 20:10 UT, $\sim$40 minutes after the start of the interplanetary type III radio bursts. 
Following the method discussed in \citet{Miteva2014}, we use a straight slope to fit the logarithm of the 60 keV electron flux profile shown in Figure~\ref{fig: wind observation}. The intersection of the fitted slope with the average pre-event background is used as the estimated earliest onset time of the energetic electron event, or 19:53:49 UT. In addition, we also find the flux that corresponds to three times the standard deviation of the pre-event background flux (or the $3\sigma$ level), shown as the horizontal dashed lines. The intersection between the same slope and the $3\sigma$ level is used to be our estimate for the latest onset time, found to be 20:03:06 UT. For the 40 keV profile, the rise can not be described by a single slope. In this case, we simply use the $3\sigma$ threshold above the background flux to identify the onset time, marked as a black circle in Figure~\ref{fig: wind observation}(a).

\begin{figure*}[!ht]
    \center
    \includegraphics[width=0.7\textwidth]{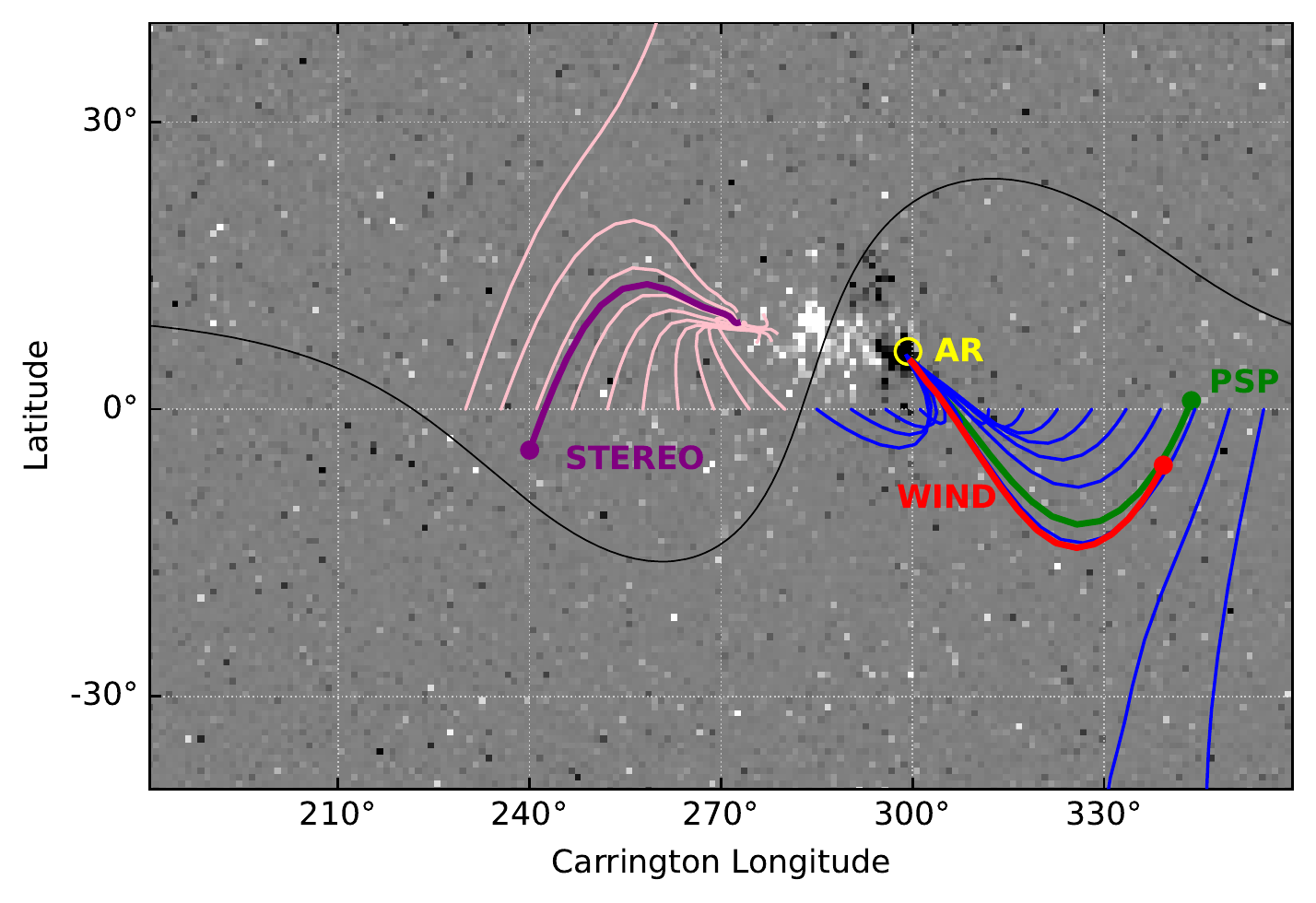}
    \caption{PFSS model mapped to the synoptic HMI magnetogram for Carrington Rotation 2216. The blue, red, and purple dots denote the location of the PSP, WIND, and STEREO-A spacecraft ballistically mapped from 0.37 AU, 1 AU, and 1 AU, respectively, to the source surface at 2.5 $R_{\odot}$ following the Parker spiral. They are all magnetically connected to AR 12738, as shown by the PSFF field lines (green, red, and purple curves). However, while both WIND and PSP are well connected to the jet region (marked by the yellow open circle), STEREO-A is connected to the opposite side of the AR far away from the jet region. The black solid line is the source surface magnetic neutral line mapped to the solar surface.}.
    \label{fig: connectivity}
\end{figure*}

If assuming scatter-free propagation along the Parker spiral with a path length of 1.2 AU (The green line shown in Figure~\ref{fig: spacecraft location}(a)), the release time of the 60 keV electrons near the solar surface would be $\sim$19:33 UT,  which is delayed by $\sim$11 minutes after the onset of the microflare/type III radio bursts at $\sim$19:22 UT (19:30:37 UT as observed from 1 AU, corrected with the light traveling time of 8.3 minutes). If we assume the electrons detected by WIND are released at the same time as the onset of the interplanetary type III radio burst event, the estimated path length for 60 keV electrons is approximately 1.68 AU, which is greater than the expected distance of 1.2 AU along the Parker spiral. Such a delay has been frequently observed in the past for impulsive SEE events \citep[e.g.,][]{Haggerty2002, Cane2003a, Kahler2006, Kahler2007, Klassen2012}, and has been interpreted as either a delayed injection through, e.g., acceleration by coronal shocks developed in the upper corona \citep{Haggerty2002} and/or effects due to transport, which increase the effective path length of the energetic electrons \citep{Cane2003a}. 

STEREO-A also carries an energetic particle instrument: the Solar Electron and Proton Telescope (SEPT; \citealt{Muller2008}). However, it did not detect any enhancement in the measured energetic electron flux in the 30--400 keV range. As we will discuss in the next subsection,  the magnetic field line connecting to STEREO-A is rooted at the opposite side of the active region far away from the jet location, which might explain the non-detection of energetic electrons arriving at the spacecraft.

\subsection{The Solar Source: Nonthermal Electrons Escaped from a Solar Jet} \label{sec:Magnetic connectivity}

In Section \ref{sec:directivity}, we have shown that the source trajectory of the interplanetary type III radio burst event likely follows magnetic field lines in interplanetary space that are generally connected to PSP at 0.37 AU and WIND (Earth) at 1 AU. Then, in Section \ref{sec:in situ}, we have shown that both PSP and WIND observed \textit{in situ} signatures of energetic electrons that arrived at the spacecraft, while STEREO-A did not. Here we further discuss the solar source of the type-III-burst-emitting energetic electrons by examining the magnetic connectivity of the three spacecraft and concurrent microwave imaging spectroscopy observations of the solar jet/microflare event. 

\begin{figure*}[ht!]
    \center
    \includegraphics[width=1.0\textwidth]{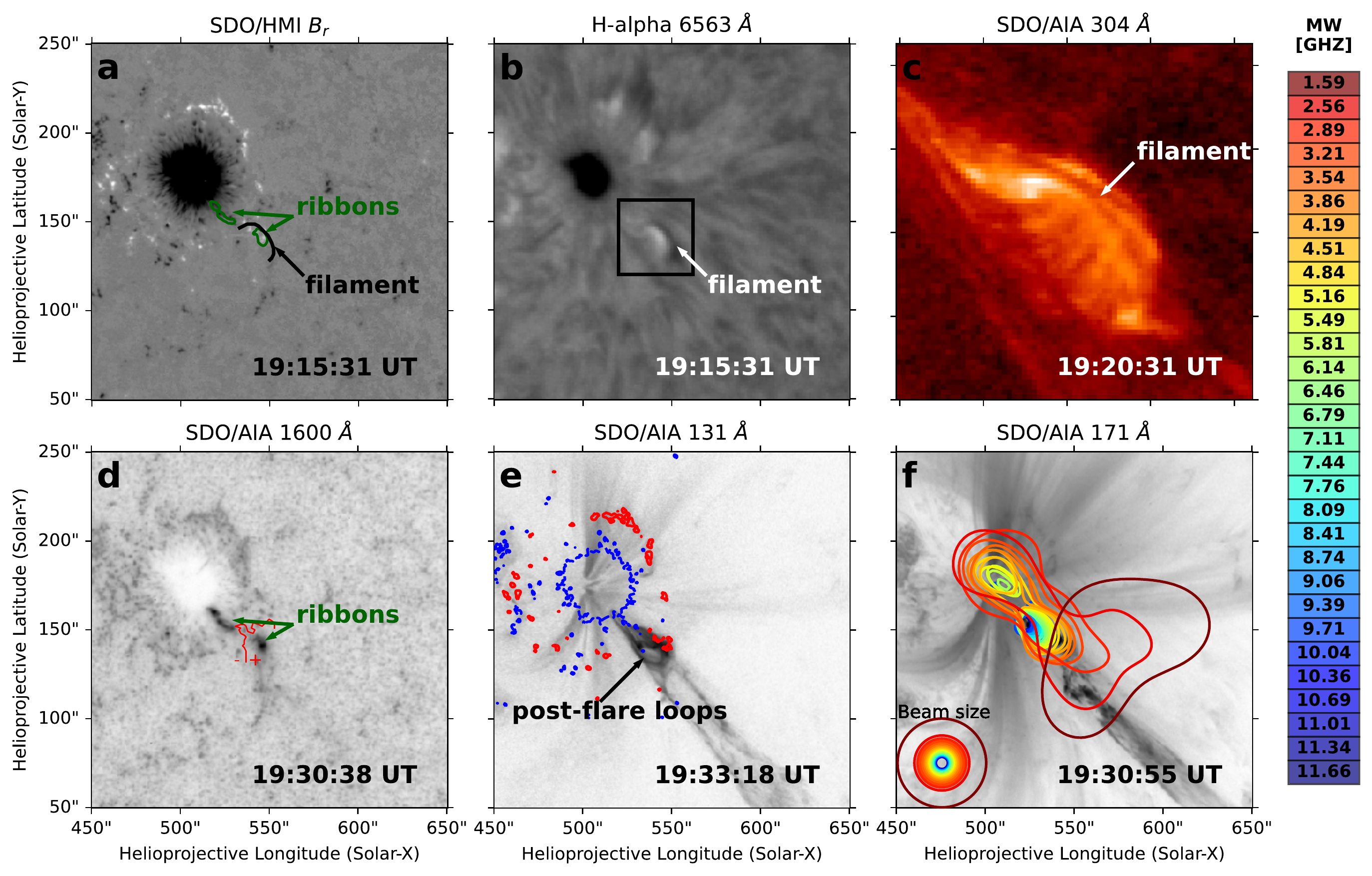}
    \caption{(a) SDO/HMI radial field magnetogram of the main sunspot. The black curve corresponds to the dark filament seen in the H$\alpha$ image shown in (b) and the SDO/AIA 304 \AA\ image in (c). (d) SDO/AIA 1600 \AA\ image shown in reversed color scale (black is bright). Two ribbons are marked, which are also outlined as green contours in (a). The red curve represents the magnetic polarity inversion line. (e) SDO/AIA 131 \AA\ image shown in reversed color scale. Blue and red contours show radial magnetic field of $-$500 G and 300 G respectively. (f) EOVSA microwave source at multiple frequencies (50$\%$ contours) overlaid on SDO/AIA 171 \AA\ image. They are colored from red in blue according to the color bar on the right. The FWHM beam sizes at different frequencies used to restore the EOVSA images are shown at the bottom left corner with the same color scheme.}
    \label{fig:filament}
\end{figure*}

\begin{figure*}[ht!]
    \center
    \includegraphics[width=1.0\textwidth]{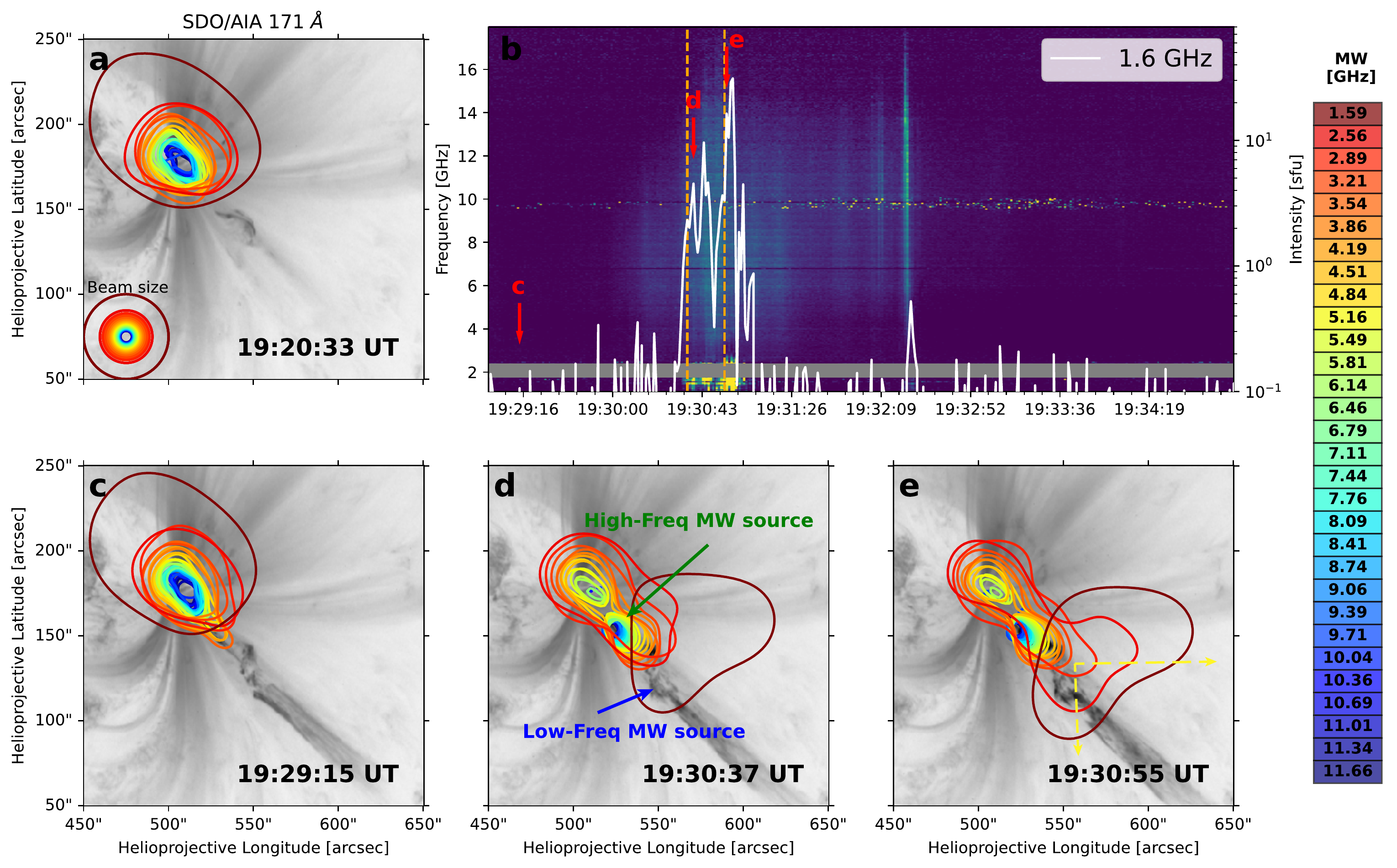}
    \caption{Evolution of the multi-frequency EOVSA microwave source from 1.6 GHz to 12 GHz, shown as 50$\%$ contours colored from red to blue for (a) pre-flare phase and (c)--(e) during the peak time. All the background is SDO/AIA 171 \AA\ image. The corresponding times in (c)--(e) are shown as red arrows in the cross-power microwave dynamic spectrum in (b). The white curve shows the 1.6 GHz light curve. The two orange dashed lines in (b) show the initiation times of the interplanetary type III radio bursts discussed in Section~\ref{sec:freq_drift}. The EOVSA beam sizes are shown at the bottom left corner in (a).}
    \label{fig:microwave_evolv}
\end{figure*}

First, we investigate the solar source region of the \textit{in situ} SEE event observed by PSP and WIND. We follow the widely-used PFSS extrapolation method \citep[see, e.g.,][]{Stansby2020,Gieseler2022} to first ballistically map the spacecraft locations following the Parker spiral to the source surface where the field lines are predominantly radial (taken to be at a heliocentric distance of 2.5$R_{\odot}$). Then, we trace the field lines back to the solar surface using the PFSS-extrapolated potential field model based on the same SDO/HMI synoptic map discussed in Section \ref{sec:overview}. The solar wind speed is assumed to be 350 km~s$^{-1}$. Figure \ref{fig: connectivity} shows our results.
The green, red, and purple circles mark the locations of the PSP, WIND, and STEREO-A spacecraft ballistically mapped using the nominal Parker spiral from the spacecraft's respective location to the source surface at 2.5$R_{\odot}$. As shown in Figure~\ref{fig: connectivity}, PSP and WIND are both well connected magnetically to AR 12738 where the jet occurs. Meanwhile, STEREO-A is connected to the remote northeast edge of the same active region with a positive polarity. The lack of magnetic connectivity between the STEREO-A and the jet region explains the non-detection of energetic electrons by STEREO-A. In addition to the temporal coincidence discussed in Section~\ref{sec:freq_drift}, the PFSS magnetic modeling results reaffirm that the solar source of the interplanetary type III burst event and the associated \textit{in situ} SEE signatures did indeed originate from the active region which produced the microflare-related solar jet. As shown in Figure~\ref{fig:jet_overview}(f), the jet has a favorable magnetic geometry featuring widely open field lines toward the west, which allows the nonthermal electrons to access interplanetary space and eventually reach PSP and WIND.

Before the jet eruption at 19:15 UT, a dark mini-filament can be seen in the H$\alpha$ image taken by the Cerro Tololo station of the Global Oscillation Network Group (GONG; Figure~\ref{fig:filament}(b)). This mini-filament is located in a mixed polarity region off the southwest side of the main sunspot, shown in the HMI radial magnetogram in Figure~\ref{fig:filament}(a). The same mini-filament is also visible in the SDO/AIA 304 \AA\ image (Figure~\ref{fig:filament}(c)), which is sensitive to plasma at chromospheric temperatures. This filament starts to rise at $\sim$19:20 UT followed by the jet eruption at 19:30 UT. Meanwhile, two bright footpoints start to appear in the SDO/AIA 1600 \AA\ image (Figure~\ref{fig:filament}(d)). The footpoints reside at opposite ends of the northern leg of the mini-filament with opposite magnetic polarities (Figure~\ref{fig:filament}(a)). After the jet eruption at $\sim$19:33 UT, a bright, arcade-like loop appears in the SDO/AIA 131 \AA\ image (Figure~\ref{fig:filament}(e)), indicative of flare-heated hot plasma at $\sim$10 MK. The magnetic, H$\alpha$, and EUV signatures of the jet eruption are all broadly consistent with the mini-filament-driven jet scenario discussed in, e.g., \citet{Sterling2015}.   

At the time of the event, EOVSA was observing the Sun with 451 science spectral channels spanning the 1--18 GHz range at a 1-s time cadence. For imaging spectroscopy, we combine the fine spectral channels into 51 spectral windows, each of which has a bandwidth of 325 MHz. As the frequency range at $>$12 GHz has a weak response, in this study, we only show the results from the first 30 spectral windows ranging from 1.6 GHz to 12 GHz. The EOVSA data were first calibrated using a celestial source and then were self-calibrated and averaged to a cadence of 2 s. Images were then reconstructed using the \texttt{tclean} task available in \texttt{CASA} \citep{2007ASPC..376..127M}. A frequency-dependent circular restoring beam of $80''/\nu_{\rm GHz}$, where $\nu_{\rm GHz}$ is the frequency in GHz, is used to restore all images\footnote{The natural beam has a FWHM size of approximately $94''/\nu_{\rm GHz}$ and $62''/\nu_{\rm GHz}$ for the major and minor axis, respectively.} (color circles shown in the bottom-left corner of Figure~\ref{fig:filament}(f)). The multi-frequency EOVSA microwave sources during the peak are shown in Figure~\ref{fig:filament}(f) as color contours (red to blue indicate increasing frequency). In Figures~\ref{fig:microwave_evolv}(c)--(e), we show four snapshots of the multi-frequency EOVSA images obtained during the pre-event quiescent period (panel a), the initiation phase of the event (panel c), the first 1.6 GHz peak (panel d), and the second 1.6 GHz peak (panel e, which is identical to Figure~\ref{fig:filament}(f)), respectively.

Prior to the event, the microwave sources are located at the main sunspot (Figure~\ref{fig:microwave_evolv}(a)) likely due to gyroresonance emission from the quiescent thermal plasma \citep{Zheleznyakov1962,Kakinuma1962,Lee2007}. At the beginning of the jet/microflare event, a small extension of the microwave source along the direction of the jet material starts to be visible (Figure~\ref{fig:microwave_evolv}(c)). 
During the peaks of the event, the pre-event sunspot source is still present, but a new microwave source becomes clearly visible at the location of the jet eruption. 
The source morphology in the low and high-frequency regimes appears distinctly different. At high frequencies (above $\sim$3 GHz), the microwave source coincides very well with a bright arcade-like structure at the base of the jet spire.
In contrast, at low frequencies ($<$3 GHz) the microwave source (red contour marked by the blue arrow in Figure~\ref{fig:microwave_evolv}(d) and the first three red contours in Figure~\ref{fig:microwave_evolv}(e)) is located above the arcade. The low-frequency source appears at the first peak time of the 1.6 GHz emission at 19:30:37 UT and lasts about half a minute. It coincides with the jet peak but has a much wider angle up to about $90^{\circ}$, indicated by the yellow dashed lines in Figure~\ref{fig:microwave_evolv}(e). 

\section{Summary and Discussion} \label{sec:Summary and Discussion}

In this study, we have presented a multi-instrument, multi-perspective observation of an interplanetary type III radio burst event accompanied with \textit{in situ} energetic electrons. We then identify the source region of the event as a solar jet in the periphery of a solar active region through the analysis of temporal association and magnetic connectivity. Our observations are briefly summarized as follows:
\begin{enumerate}[topsep=2pt,itemsep=0ex,partopsep=1ex,parsep=1ex]
    \item By utilizing observations from three well-separated spacecraft (PSP/FIELDS, WIND/WAVES, and STEREO-A/WAVES), we find that the source motion and emission pattern of the interplanetary type III radio burst are generally directed toward PSP at 0.37 AU and WIND at 1 AU. 
    \item PSP/FIELDS observations of the interplanetary type III radio burst event show that the electromagnetic radiation reaches the local plasma frequency line, which suggests that the escaping electron beams have reached at least the same radial distance from the Sun as that of PSP. Meanwhile, PSP/SWEAP recorded an enhanced low-energy ($<$1 keV) suprathermal electron burst event with a beamed anti-sunward angular distribution, implicating that the electron beams have reached the close vicinity of the PSP.
    \item WIND/3DP at 1 AU recorded a weak \textit{in situ} SEE event at 40--60 keV. The onset time, which occurs about 30 minutes after the interplanetary type III event, shows a characteristic energy dependence, implying a solar origin. The inferred release time of the energetic electrons at the solar surface (assuming free propagation) is within $\sim$10 minutes of the microwave burst onset, suggesting a common source.
    \item STEREO-A/SEPT did not detect any enhancement of energetic electrons at 30--400 keV.  
    \item The onset time of the interplanetary type III radio burst coincides very well with the occurrence of a solar jet in AR 12738. PFSS extrapolation results show that PSP and WIND were magnetically connected to the jet region, but STEREO-A was not, which may explain the detection/non-detection of the energetic electrons by the three spacecraft.
    \item The solar jet occurs in an open field line region at the periphery of the AR and is accompanied by a co-located EOVSA nonthermal microwave source. In particular, the 1--2.6 GHz microwave source, which we interpret as the decimetric counterpart of the interplanetary type III radio bursts, features a morphology that subtends a large angle of 80--90$^{\circ}$ in the open field region, which is 8--9 times wider than the EUV jet spire itself ($\sim\!10^{\circ}$).   
\end{enumerate}

\begin{figure*}
    \center
    \includegraphics[width=1\textwidth]{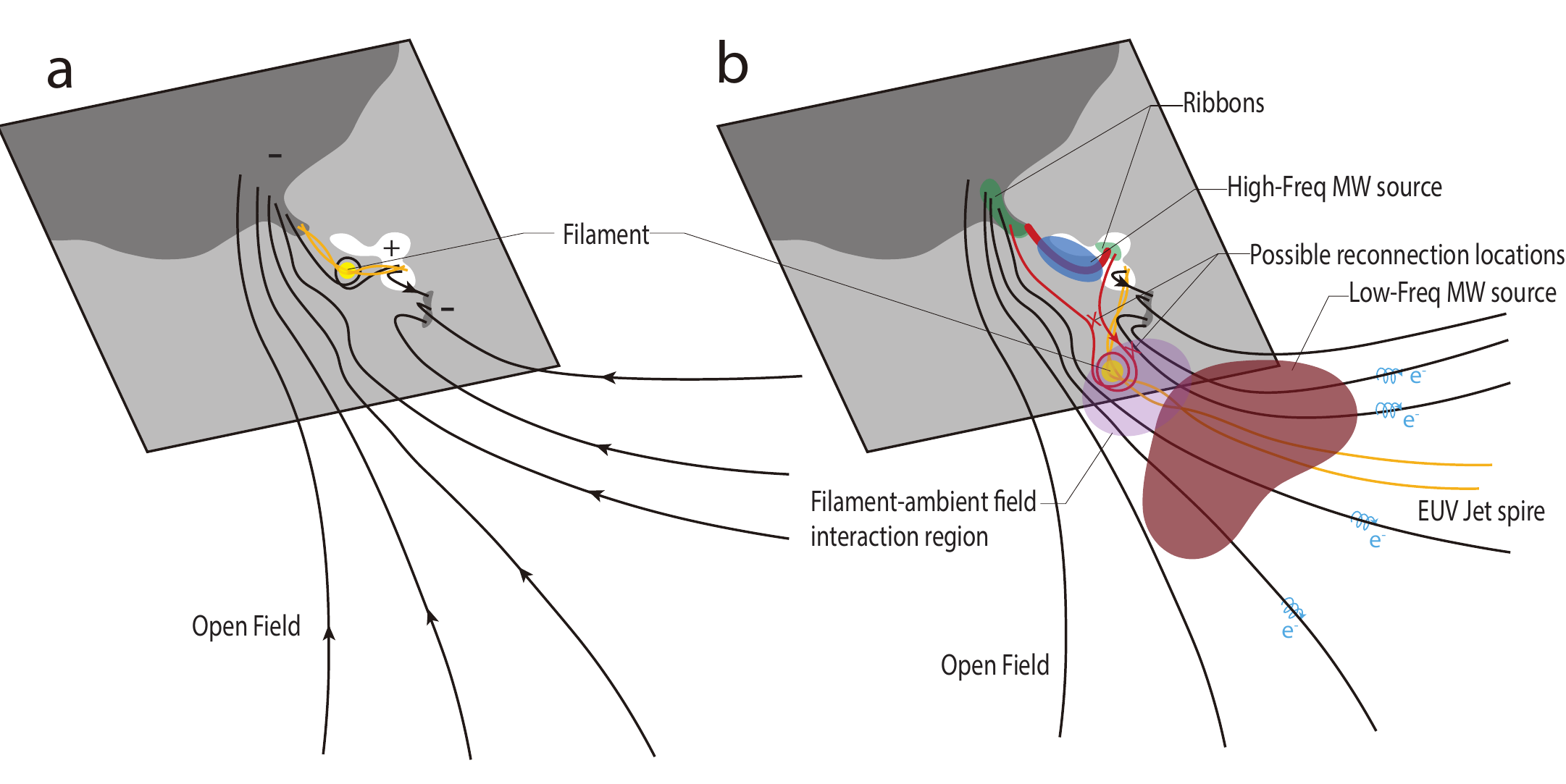}
    \caption{Schematic of the solar jet event associated with escaping energetic electrons. (a) Before the jet eruption, a mini-filament lies at the southwest side of the main sunspot where an opposite magnetic polarity is present. (b) During the jet eruption, a bright flare arcade is formed due to magnetic reconnection trailing the erupting mini-filament. Energetic electrons trapped in the arcade produce the high-frequency microwave (MW) source via incoherent gyrosynchrotron emission (blue shade). The purple shaded region represents the complex interaction region between the erupting filament and different ambient magnetic field lines. Energetic electrons entering this region could have access to these field lines and escape with a wide opening angle (see text for more discussion). As a result, a low-frequency microwave source is formed due to coherent plasma radiation (red shade), which appears to subtend a much wider angle than that of the highly collimated EUV jet spire.}
    \label{fig:reconnection}
\end{figure*}

In the following, we incorporate the multifaceted remote-sensing and \textit{in situ} observations and discuss them in a coherent picture, shown as a schematic in Figure~\ref{fig:reconnection}. The event is driven by the eruption of a mini-filament, or a small-scale magnetic flux rope, shown as the yellow feature in Figure~\ref{fig:reconnection}(a). 
As depicted by \citet{Sterling2015} and \citet{Wyper2017}, the eruption may lead to magnetic reconnection occurring at locations both trailing and above the erupting filament/flux rope, referred to inner and outer reconnection, respectively. The inner reconnection, which is akin to that in the standard CSHKP eruptive flare scenario \citep{Carmichael1964, Sturrock1966, Hirayama1974,Kopp1976}, leads to the formation of a post-reconnection flare arcade with a pair of conjugate footpoints, shown in Figure~\ref{fig:jet_overview}(e) and (d) respectively. The outer reconnection between the filament and ambient field allows the eruption to ``break out'' and untwist, forming the jet material seen in EUV. The reconnection-driven energy release produces energetic electrons. They may be accelerated at the reconnection null point(s) \citep{Drake2006,Browning2010,Stanier2012,chen2018,Che2021} or in the reconnection-driven outflows or shocks \citep{Yokoyama1996, Glesener2012,Chen2015,Kong2019}. The accelerated nonthermal electrons trapped in the post-reconnection arcade emit incoherent gyrosynchrotron radiation, observed as the compact, high-frequency ($>$3 GHz) EOVSA microwave source (depicted as the blue-shaded region). 

The upward-directed nonthermal electrons, in contrast, have access to the open field lines in the direction of the jet eruption. Thanks to their close magnetic connectivity with the jet region (c.f., Figure~\ref{fig: connectivity}), some of the electrons reach the PSP and WIND spacecraft and are recorded as \textit{in situ} energetic electron events. STEREO-A was not magnetically connected to the jet region and did not record any energetic electron event. 

The nonthermal electrons propagating along open field lines can quickly develop into a beam-like distribution with bump-on-tail instabilities and emit coherent plasma radiation near the plasma frequency or its second harmonic \citep{Lin1990,Reid2011}.  Assuming harmonic plasma radiation, the estimated plasma density of the observed 1--2.6 GHz microwave source is $3\times10^9$--$3\times10^{10}$~cm$^{-3}$, which is typical for the low corona in an active region. The escaping nonthermal electrons enter interplanetary space and generate the meter-decameter wave type III radio bursts observed by e-Callisto, PSP, WIND, and STEREO in a broad frequency range from $\sim$100 MHz down to $<$0.1 MHz. We note that each electron beam can generate a type III radio burst with a broad, dipole-like, or quadruple-like emission pattern \citep{Zheleznyakov1970}. In addition, particle and radio wave instruments have different capabilities in detecting weak signal against the background noise. Therefore, it is not surprising that STEREO-A recorded a weak type III radio burst event, despite not detecting the energetic electrons \emph{in situ}. 

We also point out that the 1.6--3 GHz EOVSA microwave source located above the post-reconnection arcade is likely different from that reported by \citet{Glesener2018}, interpreted by the authors as a source due to incoherent gyrosynchrotron radiation. Based on the spectral properties shown in Figure~\ref{fig:crosspower}, we have concluded that our 1--2.6 GHz microwave source should resemble the previously reported decimetric type III radio bursts observed by VLA in a similar frequency range (1--2 GHz L band), which had a much higher frequency and time resolution to reveal their detailed spectral, temporal, and polarization properties \citep{chen2013,chen2018}. 

One striking finding is that, contrary to most existing jet models in which nonthermal electrons are often assumed to escape along the highly collimated jet spire, the morphology of our observed 1--3 GHz microwave source is by no means collimated. In stark contrast, the source features a fan-like morphology with an opening angle of $\sim$80--90 degrees at any instant (2-s snapshot in our case), rather than the narrow EUV jet that opens by only $\sim$10 degrees. Note that, in \citet{chen2018}, with an ultra-high time resolution of 50 ms provided by the VLA, the decimetric type III bursts were resolved into a group of $>$10 individual bursts within a duration of $\sim$1 s. Although the trajectory of the individual electron beam generating each burst (in 50 ms) formed a highly collimated, nearly straight line, all the beams of the $\sim$1-s-duration type III burst group were shown to fan out to a much wider angle ($\sim\!30^{\circ}$ in their event). In addition, although they seemed to propagate along the same general direction, the electron beams did not show an exact co-alignment with the EUV jet spire. With EOVSA's limited time resolution, we are unable to resolve these individual bursts in our event. However, it remains a possibility that our case could be similar to the previously reported VLA event.

Our observations imply that the accelerated, upward-going nonthermal electrons have access to a far wider angle than that of the jet spire itself, by as much as one order of magnitude (or two orders of magnitude in solid angle). Possible causes include a wide distribution of the electron acceleration sites in the three-dimensional volume \citep{chen2018}, reconfiguration of the magnetic field due to complex interactions between the erupting filament and the ambient field \citep{Wyper2017}, and/or cross-field diffusion of the energetic electrons themselves. The wide opening angle of the escaping, type-III-burst-emitting electron beams has the following broader implications: 
\begin{enumerate}[topsep=2pt,itemsep=0ex,partopsep=1ex,parsep=1ex]
    \item Energetic electrons may be injected into a broad range of interplanetary magnetic field lines near their source at the solar surface. Therefore, the wide spread of flare events that produce observed SEE events at a single observer location \citep[e.g.,][]{Anderson1966,Lin1974,Reames1999} or SEE/SEP events that deviate significantly from their nominal magnetic field footpoints \citep{Wiedenbeck2013,Leske2020} does not necessarily have to be interpreted solely in terms of cross-field diffusion effects or secondary acceleration site(s) in the upper corona. In \citet{Lin1974}, flares associated with impulsive SEE events observed in 1964--1972 are shown to distribute in a wide longitudinal range from $\sim\!30^{\circ}$ east to $90^{\circ}$ west, with a full width of $70^{\circ}$. A more recent statistical study of $^3{He}$-rich impulsive SEE events from 1995 to 2005 observed by WIND/3DP \citep{wang2012} also suggests that the longitudinal distribution of such SEE-associated flares distribute widely from $90^{\circ}$ east to $90^{\circ}$ degrees west. Such an observed longitudinal width, which can be interpreted as the (statistically averaged) width of the injection cone of energetic electrons in a single event, is broadly consistent with our findings 
    based on direct microwave imaging spectroscopy. 
    \item When comparing the number of flare-accelerated nonthermal electrons and those that escaped into interplanetary space, care must be taken to properly account for the injection angle of the upward-going electrons. For example, if one only uses the narrow EUV jet spire for the angle estimate, the derived total upward-going electron number based on the same measured \textit{in situ} electron flux would be underestimated by as much as two orders of magnitude. Despite this, it is worth noting that the conclusion reached by several previous studies, including \citet{Lin1971}, \citet{Krucker2007}, and \citet{Wang2021} likely still holds. These studies estimated that the number of energetic electrons escaping to interplanetary space is only a small fraction of those generating HXR/microwave emission in the post-reconnection flare arcade or footpoints. These cited studies correctly based their estimates of the injection cone of the escaping energetic electrons on the width of the statistical longitudinal distribution of the SEE-generating flares rather than that of the EUV jets themselves. 
\end{enumerate}

For our event, a comparative study between the escaping \textit{in situ} energetic electrons ($>$10s of keV) and those that remain at the flare site in a manner similar to \citet{Wang2021} (and references therein) is challenging, if not impossible, due to the lack of \textit{in situ} measurements with sufficient signal-to-noise over a wide range of energies and the unavailability of hard X-ray data. However, a rough estimate using WIND/3DP data recorded at two channels (40 keV and 60 keV) and assuming a 90$^{\circ}$ opening angle of the upward-directed electrons, the estimated power of $>$25 keV nonthermal electrons at the peak time is approximately $1\times10^{21}$~ergs~s$^{-1}$. Consistent with earlier findings, the nonthermal power of upward-directed electrons is orders of magnitude lower than that of downward-directed nonthermal electrons produced in microflares inferred from the X-ray data, which has a nonthermal power of $>$25 keV electrons ranging from 10$^{22}$--10$^{27}$~ergs~s$^{-1}$ \citep{Hannah2008}. 
We are extending our investigations to larger impulsive SEE events that are well connected to their source regions on the Sun and will report our results in future studies.

\begin{acknowledgments}

The authors acknowledge Drs. Sophie Musset, Radoslav Bučík, Sung Jun Noh, and Marc Pulupa for helpful discussions. The work by MW and BC is mainly supported by NASA grant HSO Connect 80NSSC20K1282, NASA grant HSO PSP data support 80NSSC20K0026, and NSF CAREER grant AGS-1654382 to NJIT. 
SY, JL, and HW acknowledge additional support by NASA HSO Connect grant 80NSSC20K1283, NSF grants AGS-2114201, and AGS-2229064 to NJIT.
CMSC is supported in part by NASA’s Parker Solar Probe Mission, contract NNN06AA01C. 
EOVSA operations are supported by NSF grants AST-1910354 and AGS-2130832 to NJIT. Parker Solar Probe was designed, built, and is now operated by the Johns Hopkins Applied Physics Laboratory as part of NASA’s Living with a Star (LWS) program. 
\end{acknowledgments}

\vspace{5mm}
\facilities{OVRO:SA, Parker, SDO, STEREO, WIND, GONG}
\software{CASA \citep{2007ASPC..376..127M},
          Astropy \citep{2018AJ....156..123A}, 
          SunPy \citep{2020ApJ...890...68S}, pfsspy \citep{Stansby2020}, Numpy \citep{Harris2020}, spacepy \citep{morley2022}}

\bibliography{in_situ_type3}{}
\bibliographystyle{aasjournal}



\end{document}